\documentclass[oneside]{ar2e}
\usepackage{graphicx}

\newcommand {\sigmann}{\sigma^\mathrm{NN}_\mathrm{inel}}

\begin{document}
\input epsf.tex    %<-If you need EPS figures to be
                   %  called in {figure} environment for PC
%\input epsf.def   %<-If you need EPS figures to be
                   %  called in {figure} environment for Macintosh

\input psfig.sty

\jname{Annu. Rev. Nucl. Part. Sci.}
\jyear{2007}
\jvol{57}
%\ARinfo{1056-8700/97/0610-00}

\title{Glauber Modeling in High Energy \\Nuclear Collisions}

\markboth{Miller, Reygers, Sanders, Steinberg}{Glauber Modeling in Nuclear Collisions}

\author{Michael L. Miller$^1$ (mlmiller@mit.edu)\\Klaus Reygers$^2$ (reygers@ikp.uni-muenster.de)\\Stephen J. Sanders$^3$ (ssanders@ku.edu)\\Peter Steinberg$^{4,*}$ (peter.steinberg@bnl.gov)
\affiliation{$^1$ Massachusetts Institute of Technology, Cambridge, MA, $^2$ University of M{\"u}nster, Germany, $^3$University of Kansas, Lawrence, Kansas, $^4$Brookhaven National Laboratory, Upton, NY}
}
\begin{keywords}
Glauber modeling, heavy ion physics, number of participating nucleons, number of binary collisions, impact parameter, eccentricity
\end{keywords}

\newpage

\begin{abstract}
  This is a review of the theoretical background, experimental
  techniques, and phenomenology of what is called the ``Glauber
  Model'' in relativistic heavy ion physics.  This model is used to
  calculate ``geometric'' quantities, which are typically expressed as
  impact parameter ($b$), number of participating nucleons
  ($N_\mathrm{part}$) and number of binary nucleon-nucleon collisions
  ($N_\mathrm{coll}$).  A brief history of the original Glauber model
  is presented, with emphasis on its development into the purely
  classical, geometric picture that is used for present-day data
  analyses.  Distinctions are made between the ``optical limit'' and
  Monte Carlo approaches, which are often used interchangably but have
  some essential differences in particular contexts.  The methods used
  by the four RHIC experiments are compared and contrasted, although
  the end results are reassuringly similar for the various geometric
  observables.  Finally, several important RHIC measurements are
  highlighted that rely on geometric quantities, estimated from
  Glauber calculations, to draw insight from experimental observables.
  The status and future of Glauber modeling in the next generation of
  heavy ion physics studies is briefly discussed.
\end{abstract}

\maketitle

\newpage
\section{Introduction}
\label{sect:intro}
Ultra-relativistic 
collisions of nuclei produce the highest multiplicities of outgoing
particles of all subatomic systems known in the laboratory.  Thousands
of particles are created when two nuclei collide head-on, generating
event images of dramatic complexity compared with proton-proton
collisions.  The latter is a natural point of comparison, as nuclei
are themselves made up of nucleons, i.e. protons and neutrons.  Thus,
it is natural to ask just how many nucleons are involved in any
particular collision or, more reasonably, in a sample of selected
collisions.  It is also an interesting to consider other ways to
characterize the overlapping nuclei, e.g. their shape.

While this problem would seem intractable, with the femtoscopic length
scales involved precluding direct observation of the impact parameter
($b$) or number of participating nucleons ($N_\mathrm{part}$) or binary
nucleon-nucleon collisions ($N_\mathrm{coll}$), theoretical techniques have
been developed to allow estimation of these quantities from
experimental data.  These techniques, which consider the multiple-
scattering of nucleons in nuclear targets, are generally referred to
as ``Glauber Models'' after Roy Glauber.  Glauber pioneered the use of
quantum mechanical scattering theory for composite systems, describing
non trivial effects discovered in cross sections for proton-nucleus
(p+A) and nucleus-nucleus (A+B) collisions at low energies.

Over the years, a variety of methods were developed for calculating
geometric quantities relevant for p+A and A+B 
collisions~\cite{Czyz:1969jg,Glauber:1958,Bialas:1977pd}.  Moreover,
a wide variety of experimental methods were tried to connect
distributions of measured quantities to the distributions of
geometric quantities.  This review is an attempt to explain how the
RHIC experiments grappled with this problem, and largely succeeded,
both because of good-sense experimental and theoretical thinking.

Heavy ion physics entered a new era with the turn-on of the RHIC
collider in 2000~\cite{Back:2000gw}.  Previous generations of heavy ion 
experiments searching for the Quark Gluon Plasma (QGP) had
focused on particular signatures suggested by theory.  From
this strategy, experiments generally measured observables in
different regions of phase space, or focused on particular
particle types.  RHIC experiments were designed in a comprehensive
way, with certain regions of phase space being covered by multiple
experiments.  This allowed systematic cross checks of various
measurements between experiments, increasing the credibility
of the separate 
results~\cite{Arsene:2004fa,Adcox:2004mh,Back:2004je,Adams:2005dq}.  

Among the most fundamental observables shared between the
experiments were those relating to the geometry of the collision.
Identical Zero-Degree Calorimeters~\cite{Adler:2000bd} 
were installed in all experiments
to estimate the number of spectator nucleons, and all experiments
had coverage for multiplicities and energy measurements over a
wide angular range.  This allowed a set of systematic studies
of centrality in d+Au and A+A collisions, providing one of the
first truly extensive datasets all of which characterized by
geometric quantities. 

This review will cover the basic information a newcomer to the
field should know to understand how centrality is estimated by
the experiments, and how this is related to nucleus-nucleus collisions.
Section \ref{sect:theory} will discuss the history of the Glauber model
by reference to the theoretical description of nucleus-nucleus
collisions.
Section \ref{sect:experiment} will discuss how experiments measure
centrality by a variety of methods and relate that by a 
simple procedure to geometrical quantities.
Section \ref{sect:physics} will illustrate the relevance of various
geometrical quantities by reference to actual RHIC data.  These
examples will illustrate how a precise quantitative grasp of the
geometry allows qualitatively new insights into the dynamics of
nucleus-nucleus collisions.
Finally, section \ref{sect:discussion} will assess the current state
of knowledge and point to new directions in our understanding
of nuclear geometry and its impact on present and future
measurements.

\newpage
\section{Theoretical foundations of Glauber modeling}
\label{sect:theory}
\subsection{A brief history of the Glauber model}
The Glauber model was developed to address the problem of high energy
scattering with composite particles.  This was of great interest to
both nuclear and particle physicists, who have both benefited from the
insights of Glauber in the 1950's.  In his 1958 lectures, Glauber
presented his first collection of various papers and unpublished work
from the 1950's \cite{Glauber:1958}.  Some of this was updated work
started by Moliere in the 1940's, but much of it was in direct
response to the new experiments involving protons, deuterons and light
nuclei.  Up to that point, there had been no systematic calculations
treating the many-body nuclear system either as a projectile or
target.  Glauber's work put the quantum theory of collisions of
composite objects on a firm basis, and provided a consistent
description of experimental data for protons incident on deuterons and
larger nuclei \cite{Glauber:1955qq, Franco:1965wi}.  Most striking were
the observed dips in the elastic peaks whose position and magnitude
were predicted by Glauber's theory by Czyz and Lesniak in
1967~\cite{Czyz:1967}.

It was only in the 1970's when high energy beams of hadrons and nuclei
were systematically scattered off of nuclear targets.  Glauber's work
was found to have utility in describing total cross sections, for
which ``factorization'' relationships were found (e.g. $\sigma_{AB}^2
\sim \sigma_{AA}\sigma_{BB}$) \cite{Fishbane:1974zi,Franco:1974gc}.
Maximon and Czyz applied the theory in its most complete form to p+A
and A+B collisions in 1969, focusing mainly on elastic collisions
\cite{Czyz:1969jg}.  Finally, Bialas, Bleszynski, and Czyz
\cite{Bialas:1976ed, Bialas:1977pd} applied Glauber's approach to
inelastic nuclear collisions, after they had already applied their
``wounded nucleon model'' to hadron-nucleus collisions This is
essentially the bare framework of the traditional ``Glauber Model'',
with all of the quantum mechanics reduced to its simplest
form~\cite{Glauber:2006gd}.  The main remaining feature of the
original Glauber calculations is the ``optical limit'', used to make
the full multiple scattering integral numerically tractable.

The approach of Bialas {\it et al.}~\cite{Bialas:1976ed} 
introduced the basic functions used in
more modern language, including the thickness function and a prototype
of the nuclear overlap function $T_{AB}$.  This paper also introduced the
convention of using the optical limit for analytic and numerical
calculations, despite full knowledge that the ``real'' Glauber
calculation is an $2(A+B+1)$-dimensional integral over the impact
parameter and each of the nucleon positions.  A similar convention
exists in most theoretical calculations of geometrical
parameters to this day.

%In many cases, especially at high energies, it is necessary
%to include the interplay between the number ``wounded nucleons'',
%generally known as ``participants'' (but not always, see NA49)
%and the number of binary collisions.  
%The relevance of the
%number of participants was elucidated in p+A collisions by
%the observation that $n_{ch} \propto N_{part}$, and assumed
%to be relevant in A+A collisions.  The physical picture 
%is that of exciting the nucleon, a strongly-bound composite
%object, to a high mass state that decays into various
%hadrons, many of which themselves cascade to lower-mass states. 
%Binary collisions, on the other hand, were thought to be the
%relevant variable for ``hard'', pointlike processes amenable
%to perturbation theory calculations.

With the advent of desktop computers,
the ``Glauber Monte Carlo'' (GMC) approach emerged as a natural
framework for use by more realistic particle production codes
\cite{Ludlam:1986dy,Shor:1988vk}.
The idea was to model the nucleus in the simplest way, as 
uncorrelated nucleons sampled from measured density
distributions.  Two nuclei could be arranged with a random
impact parameter $b$ and projected onto the x-y plane.  Then
interaction probabilities could be applied by using the
relative distance between two nucleon centroids as a stand-in
for the measured nucleon-nucleon inelastic cross section.

The GMC approach was first applied to high energy heavy ion
collisions in the HIJET model~\cite{Ludlam:1986dy} 
and has shown up in practically
all A+A simulation codes.  This includes HIJING~\cite{hijing}, 
VENUS~\cite{Werner:1988jr}, RQMD~\cite{Sorge:1989vt},
and all models which require specific production points for
the different sub-processes in a nucleus-nucleus collision,
rather than just aggregate values.

%But even with regard to sample averages (e.g. the average number
%of participants and collisions),
%it generally seemed that the choice between a semi-analytic
%optical-limit calculation and a Glauber Monte Carlo calculation
%was a matter of taste and available computing power.  However, recent 
%investigations have found that the two are qualitatively quite
%different.  The main difference is the optical limit ignoring
%the important role of event-by-event fluctuations in the distributions 
%of nucleons. This will be discussed in detail in subsection~\ref{sect:optical}

\subsection{Inputs to Glauber Calculations}
\begin{figure}[tbp]
\begin{center}
\includegraphics[width=120mm]{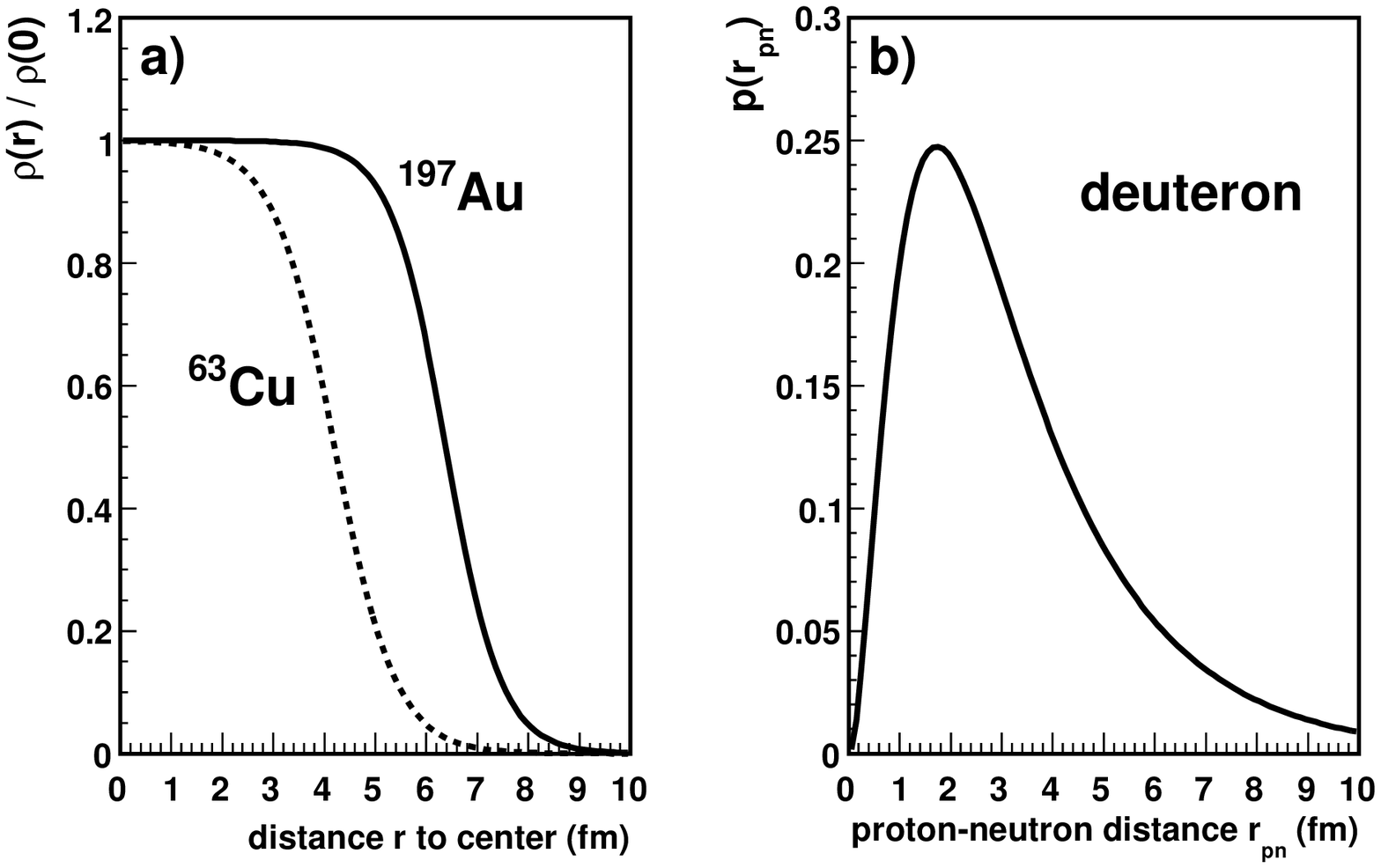}
\caption{
Density distributions for nuclei used at RHIC (a) and distribution 
of the proton-neutron distance in the deuteron as given by the 
Hulthen wave function (b). 
\label{fig:nuclear_density}
}
\end{center}
\end{figure}
In all calculations of geometric parameters using a Glauber
approach, some experimental data must be given as model inputs. 
The two most important are the nuclear charge densities measured
in low-energy electron scattering experiments and the 
energy dependence of the inelastic nucleon-nucleon cross section.
\subsubsection{Nuclear Charge Densities}
The nucleon density is usually parameterized by a Fermi
distribution with three parameters:
\begin{equation}
%\rho (r) = \rho_0 \cdot \frac{1+\frac{wr^2}{R^2}}{1 + \exp \left( \frac{r-R}{a} \right) }
\rho (r) = \rho_0 \cdot \frac{1 + w (r/R)^2}{1 + \exp \left( \frac{r-R}{a} \right) }
\end{equation}
where $\rho_0$ corresponds to the nucleon density in the center of the
nucleus, $R$ corresponds to the nuclear radius, $a$ to the ``skin depth''
and $w$ characterizes deviations from a spherical shape.  
For $^{197}$Au ($R=6.38$~fm; $a=0.535$~fm; $w=0$) and $^{63}$Cu 
($R=4.20641$~fm; $a=0.5977$~fm; $w=0$), the nuclei so far employed at RHIC, 
$\rho(r)/\rho_0$ is shown in Fig.~\ref{fig:nuclear_density}a with the
Fermi distribution parameters as given in 
ref.~\cite{landolt,DeJager:1987qc}. In the Monte Carlo procedure the radius
of a nucleon is randomly drawn from the distribution $4 \pi r^2 \,
\rho(r)$ (where the absolute normalization is of course irrelevant).
%***

%\begin{table}%
%\def~{\hphantom{0}}
%\begin{center}
%\caption{Fermi distribution parameters}\label{tab:ws}
%\begin{tabular}{@{}ccc@{}}%
%\toprule
%nucleus & $R$~(fm) & $a$~(fm)\\
%\colrule
%$^{197}$Au      & 6.38  & 0.535 \\ 
%$^{63}$Cu       & 4.20641 & 0.5977\\
%\botrule
%\end{tabular}
%\end{center}
%\end{table}

At RHIC, effects of cold nuclear matter have been studied with the aid
of d+Au collisions. In the Monte Carlo calculations the deuteron wave
function was represented by the Hulth{\'e}n form
\cite{hulthen,hodgson}
\begin{equation}
  \phi(r_\mathrm{pn}) = \frac{1}{\sqrt{2 \pi}} \frac{\sqrt{a b (a+b)}}{b-a} 
  \cdot \frac{e^{-a \, r_\mathrm{pn}} - e^{-b \, r_\mathrm{pn}}}{r_\mathrm{pn}}
  \label{eq:hulthen}
\end{equation}
with parameters $a = 0.228$\,fm$^{-1}$ and $b = 1.18$\,fm$^{-1}$
\cite{Adler:2003ii}.  The variable $r_\mathrm{pn}$ in
Eq.~\ref{eq:hulthen} denotes the distance between the proton and the
neutron. Accordingly, $r_\mathrm{pn}$ was drawn from the distribution
$p(r_\mathrm{pn}) = 4 \pi r_\mathrm{pn}^2 \phi^2(r_\mathrm{pn})$ which
is shown in Fig.~\ref{fig:nuclear_density}b.

\begin{figure}[tbp]
\begin{center}
\includegraphics[width=120mm]{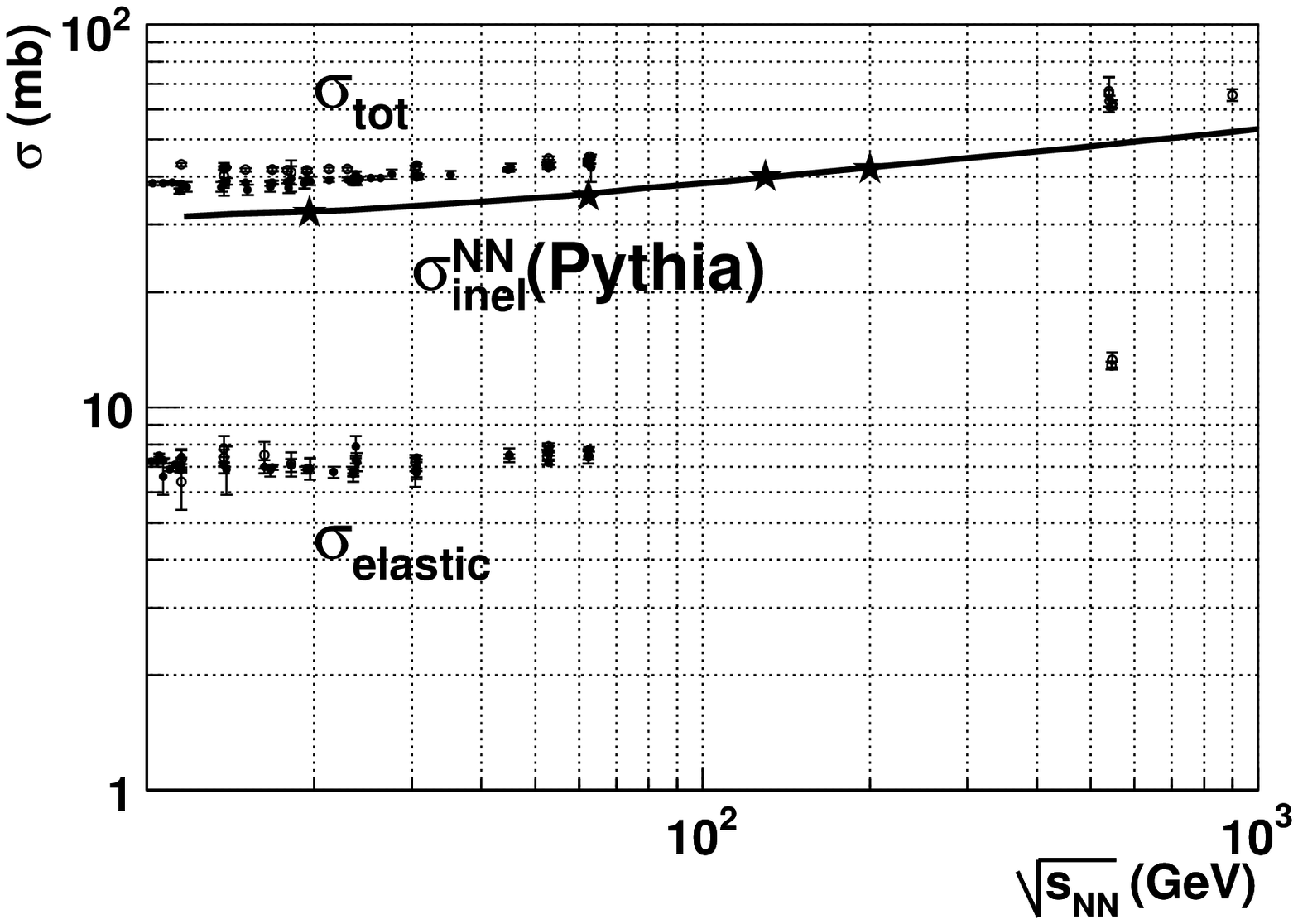}
\caption{ Inelastic nucleon-nucleon cross section
  $\sigma_\mathrm{inel}^\mathrm{NN}$ as parameterized by Pythia
  \cite{guillaud, Sjostrand:2006za} (solid line) along with data on
  total and elastic nucleon-nucleon cross sections as function of
  $\sqrt{s}$ \cite{Yao:2006px}. The stars indicate nucleon-nucleon
  cross section used for Glauber Monte Carlo calculations at RHIC
  ($\sigma_\mathrm{inel}^\mathrm{NN} = $~32.3, 35.6, 40, and 42 mb at
  $\sqrt{s_\mathrm{NN}} = $~19.6, 62.4, 130, and 200 GeV)
\label{fig:sigInelNN}
}
\end{center}
\end{figure}

\subsubsection{Inelastic Nucleon-Nucleon Cross Section ($\sigmann$)}

In the context of high energy nuclear collisions, we are typically
interested in multiparticle nucleon-nucleon processes.  As the
cross section involves processes with low momentum transfer, it
is impossible to calculate this using perturbative QCD.  Thus,
the measured inelastic nucleon-nucleon cross section ($\sigmann$)
is used as an input, and provides the only non-trivial beam-energy
dependence for Glauber calculations.
From $\sqrt{s_\mathrm{NN}} \sim 20$~GeV
(CERN SPS) to $\sqrt{s_\mathrm{NN}} = 200$~GeV (RHIC)
$\sigma^\mathrm{NN}_\mathrm{inel}$ increases from $\sim 32$~mb to
$\sim 42$~mb as shown in Fig.~\ref{fig:sigInelNN}.
Diffractive and elastic processes, which are typically
ignored in high energy multiparticle nuclear collisions,
are generally active out to
large impact parameters, and thus require full quantum mechanical
wave functions.

\subsection{Optical-limit Approximation}
\label{sect:optical}
The Glauber Model views the collision of two nuclei in terms of the
individual interactions of the constituent nucleons (see,
e.g., Ref. \cite{Wong:1995jf}). In the optical limit, the
overall phase shift of the incoming wave is taken as a sum over 
all possible two-nucleon (complex) phase shifts, with the imaginary
part of the phase shifts related to the nucleon-nucleon scattering
cross section through the optical theorem\cite{Chauvin:1983,Wibig:1998by}.    
The model assumes that at 
sufficiently high energies, these nucleons will carry sufficient momentum 
that they will be essentially undeflected as the nuclei pass through each 
other. It is also assumed that the nucleons move independently in the 
nucleus and that the size of the nucleus is large compared to
the extent of the nucleon-nucleon force.   
The hypothesis of independent linear trajectories of the constituent 
nucleons makes it possible to develop simple analytic expressions for 
the nucleus-nucleus interaction cross section and for the number of 
interacting nucleons and the number of nucleon-nucleon collisions
in terms of the basic  nucleon-nucleon cross section.

\begin{figure}[tbp]
\begin{center}
\includegraphics[width=120mm]{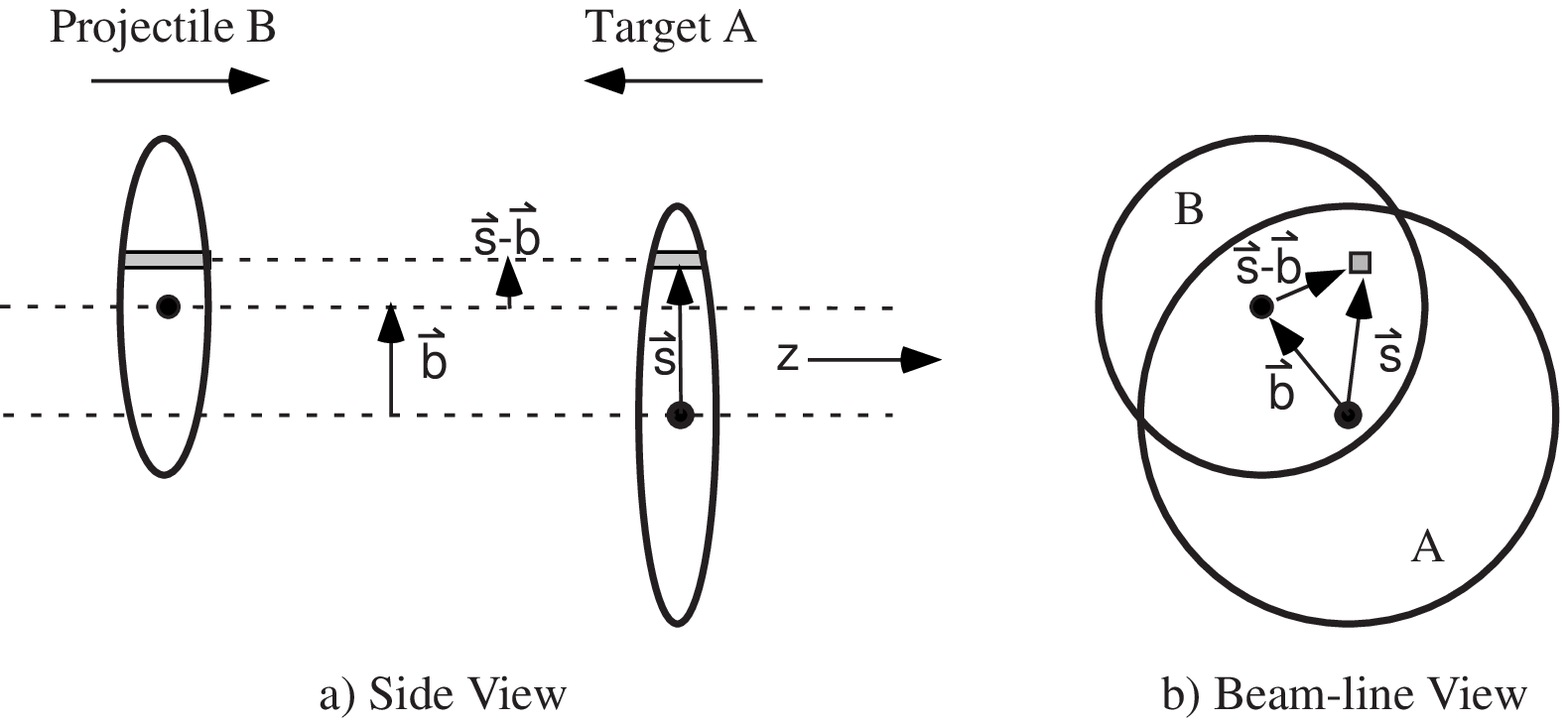}
\caption{
Schematic representation of the Optical Glauber Model geometry, with
transverse (a) and longitudinal (b) views. 
\label{fig:GlauberGeo}
}
\end{center}
\end{figure}

Consider Fig.~\ref{fig:GlauberGeo}.   Two heavy-ions, 
``target'' A and ``projectile'' B are shown colliding at relativistic 
speeds with impact parameter 
$\bf{b} $ (for colliding beam experiments the distinction between 
the target and projectile nuclei is a matter of convenience).   We focus 
on the two flux tubes located at a displacement 
$\bf{s} $ with respect to the center of the target nucleus and a 
distance $\bf{s}  - \bf{b} $ from the center 
of the projectile.   During the collision these tubes overlap. 
The probability per unit transverse area of a given nucleon being 
located in the target flux tube is 
$\hat{T}_A \left(\bf{s} \right) = \int {\hat{\rho} _A ({\bf s} ,z_A )dz_A } $, 
where 
$\hat{\rho} _A \left({\bf s} ,z_A  \right)$ is the probability per 
unit volume, normalized to unity, for finding the nucleon at location 
$\left({\bf s} ,z_A  \right)$. 
A similar expression follows for the 
projectile nucleon. The product $\hat{T}_A \left({\bf s} \right)\hat{T}_B 
\left(\bf{s}  - \bf{b} \right)d^2 s$ then gives the joint 
probability per unit area of nucleons being located in the respective 
overlapping target and projectile flux tubes of differential area $d^2 s$.  
Integrating this product over all values of $\bf{s} $  
defines the ``thickness function'' 
$\hat{T}\left(\bf{b} \right)$, with 
\begin{equation}
\hat{T}_{AB}\left(\bf{b}  \right) = \int {\hat{T}_A \left(\bf{s}  
\right)\hat{T}_B \left(\bf{s}  - \bf{b}  \right)} d^2 s.
\end{equation}

Notice that $\hat{T}\left(\bf{b}\right)$ has the unit of inverse area.  
We can interpret this as the effective overlap area for which a specific 
nucleon in A can interact with a given nucleon in B.  The probability of 
an interaction occurring is then $\hat{T}\left(\bf{b}  
\right)\sigma^\mathrm{NN}_\mathrm{inel} $, 
where  $\sigma^\mathrm{NN}_\mathrm{inel} $ is the nucleon-nucleon inelastic cross section. 
Elastic processes lead to very little energy 
loss and are consequently not considered in the Glauber-model calculations. 
Once the probably of a given nucleon-nucleon interaction has been found, 
the probably of having n such interactions between nucleus A (with $A$ 
nucleons) and B (with $B$ nucleons) is given as a binomial distribution,
\begin{equation}
P\left( n,\bf{b}  \right) = \left( {\matrix{   {AB}  \cr    n  
\cr  } } \right)\left[ \hat{T}_{AB}\left( 
\bf{b}  \right)\sigma^\mathrm{NN}_\mathrm{inel} \right]^n \left[ {1 - \hat{T}_{AB}\left( 
\bf{b}  \right)\sigma^\mathrm{NN}_\mathrm{inel} } \right]^{AB - n} 
\end{equation}
where the first term is the number of combinations for finding $n$ collisions 
out of $AB$ possible nucleon-nucleon interactions, the second term the 
probability for having exactly $n$ collisions, and the last term is the 
probability of exactly $AB-n$ misses. 

Based on this probability distribution, 
a number of useful reactions quantities can be found.   The total 
probability of an interaction between A and B is 
\begin{equation}
{{d^{2}\sigma^\mathrm{A+B}_\mathrm{inel} } 
\over {db^{2}} } \equiv p^\mathrm{A+B}_\mathrm{inel}(b) = \sum\limits_{n = 1}^{AB} {P\left( {n,
\mathord{\buildrel{\lower3pt\hbox{$\scriptscriptstyle\rightharpoonup$}}
\over  b} } \right)}  = 1 - \left[ {1 - \hat{T}_{AB}\left( \bf{b}  \right)
\sigma^\mathrm{NN}_\mathrm{inel} } \right]^{AB} .
\label{eq:p_int_AB}
\end{equation}
The vector impact parameter can be replaced by a scalar distance
if the nuclei are not polarized.  In this case, 
the total cross section can be found as 
\begin{equation}
\label{eq:sigma_AB}
\sigma^\mathrm{A+B}_\mathrm{inel}  = \int\limits_0^\infty  
{2\pi bdb\left\{ {1 - \left[ {1 - 
\hat{T}_{AB}\left( b \right)\sigma^\mathrm{NN}_\mathrm{inel} } \right]^{AB} } 
\right\}}
\end{equation}

The total number of nucleon-nucleon collisions is 
\begin{equation}
\label{eq:ncoll}
N_\mathrm{coll} \left( b \right)  = \sum
\limits_{n = 1}^{AB} {nP\left( {n,b} \right) = AB\hat{T}_{AB}\left( b \right)
\sigma^\mathrm{NN}_\mathrm{inel} }
\end{equation}
using the result for the mean of a binomial distribution.   
The number of nucleons in the target and projectile nuclei that interact 
is called either the ``number of participants'' or the ``number of wounded 
nucleons''.  The number of participants (or wounded nucleons) at 
impact parameter b is given by \cite{Bialas:1976ed,Kharzeev:1996yx}
\begin{eqnarray}
N_\mathrm{part} \left( \bf{b}  \right) & = & A
\int {\hat{T}_A \left( \bf{s}  \right)\left\{ 
{1 - \left[ {1 - \hat{T}_B \left( {
\bf{s}  - {\bf b} } \right)\sigma^\mathrm{NN}_\mathrm{inel} } \right]^B } \right\}d^2 s 
+ } \nonumber \nonumber \\
& & B\int {\hat{T}_B \left( {\bf{s}  - \bf{b} } \right)\left\{ {1 - \left[ 
{1 - \hat{T}_A \left( {\bf{ s} } \right)\sigma^\mathrm{NN}_\mathrm{inel} } \right]^A } 
\right\}d^2 s} ,
\end{eqnarray}
where it can be noted that the integral over the bracketed terms give the 
respective inelastic cross sections for nucleon-nucleus collisions:
\begin{equation}
\sigma^\mathrm{A\left( B \right)}_\mathrm{inel}  = \int {d^2 s\left\{ {1 - \left[ {1 - 
\sigma^\mathrm{NN}_\mathrm{inel} \hat{T}_{A(B)} \left( {\bf{s} } \right)} \right]^{A(B)} } 
\right\}}.
\end{equation}

The optical form of the Glauber theory is 
based on continuous nucleon density distributions.  The theory does not 
locate nucleons at specific spatial coordinates, as is the case for the 
Monte Carlo formulation that is discussed in the next section. This difference
between the optical and Monte Carlo approaches can lead to subtle differences
in calculated results, as will be discussed below.
%steve_end

\subsection{Glauber Monte Carlo approach}
%--- klaus_begin:mc_approach_v1 
%MLM --
\begin{figure}[tbp]
\begin{center}
\includegraphics[width=120mm]{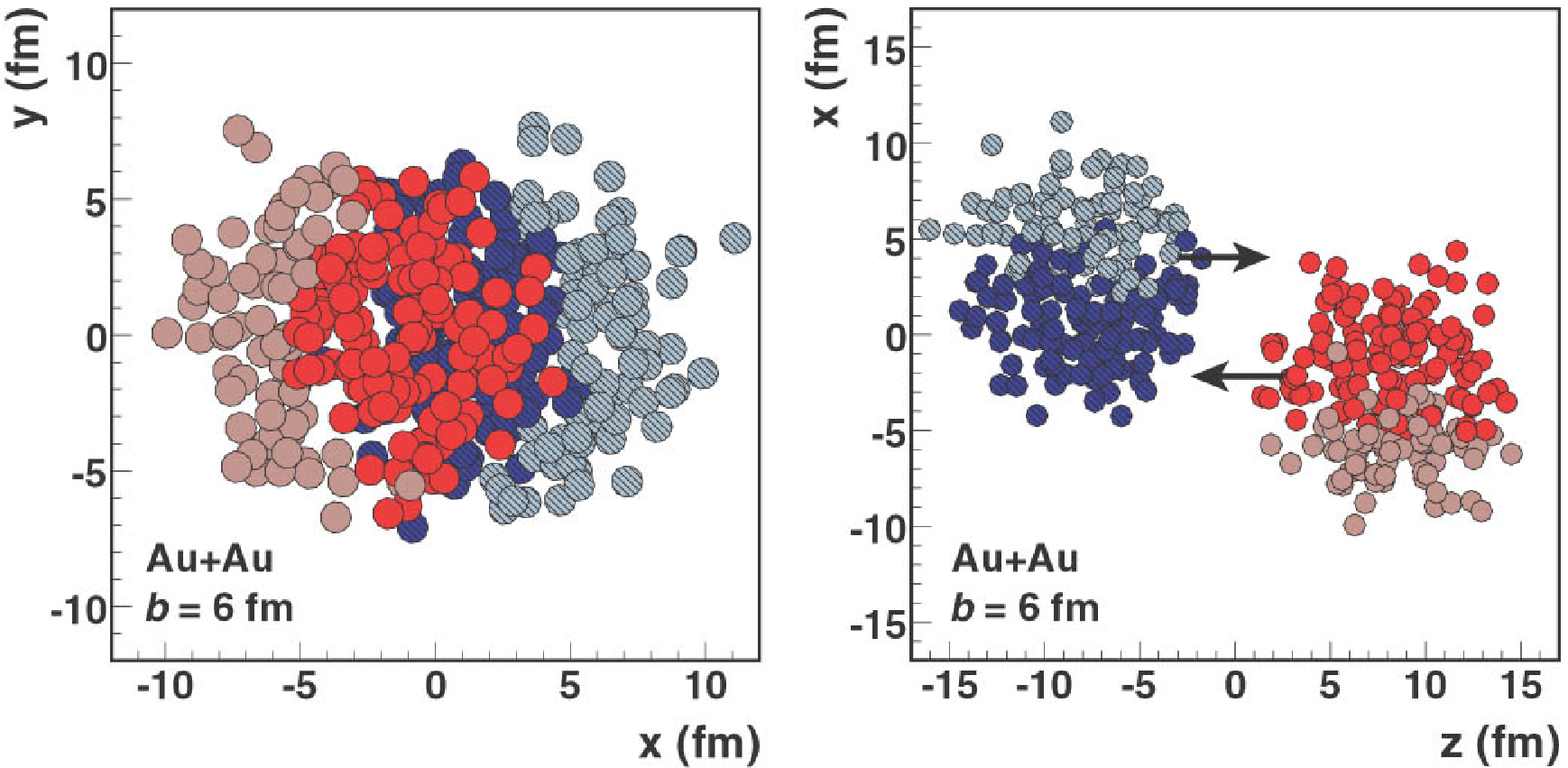}
\caption{
Glauber Monte Carlo event (Au+Au at $\sqrt{s_\mathrm{NN}} = 200$~GeV 
with impact parameter $b = 6$~fm) viewed in the transverse plane 
(left panel) and along the beam axis (right panel). The nucleons are 
drawn with a radius $\sqrt{\sigma^\mathrm{NN}_\mathrm{inel}/\pi}/2$. 
Darker disks represent participating nucleons.
\label{fig:glauber_mc_event}
}
\end{center}
\end{figure}
The virtue of the Monte Carlo approach for the calculation of geometry
related quantities like $\langle N_\mathrm{part} \rangle$ and $\langle
N_\mathrm{coll} \rangle$ is its simplicity. Moreover, it is possible
to simulate experimentally observable quantities like the charged
particle multiplicity and to apply similar centrality cuts as in the
analysis of real data. In the Monte Carlo ansatz the two colliding
nuclei are assembled in the computer by distributing the $A$ nucleons
of nucleus A and $B$ nucleons of nucleons B in three-dimensional
coordinate system according to the respective nuclear density
distribution.  A random impact parameter $b$ is then drawn from the
distribution $\mathrm{d}\sigma/\mathrm{d}b = 2 \pi b$. A
nucleus-nucleus collision is treated as a sequence of independent
binary nucleon-nucleon collisions, {\it i.e.}, the nucleons travel on
straight-line trajectories and the inelastic nucleon-nucleon
cross-section is assumed to be independent of the number of collisions
a nucleon underwent before. In the simplest version of the Monte Carlo
approach a nucleon-nucleon collision takes place if their distance $d$
in the plane orthogonal to the beam axis satisfies
\begin{equation}
d \le \sqrt{\sigma^\mathrm{NN}_\mathrm{inel}/\pi}
\end{equation}
where $\sigma^\mathrm{NN}_\mathrm{inel}$ is the total inelastic
nucleon-nucleon cross-section. 
As an alternative to
the black-disk nucleon-nucleon overlap function, {\it e.g.}, a
Gaussian overlap function can be used \cite{Pi:1992ug}. 

An illustration of a Glauber Monte Carlo event for a Au+Au collision with
impact parameter $b = 6$~fm is shown in
Fig.~\ref{fig:glauber_mc_event}.  The average number of participating
nucleons and binary nucleon-nucleon collisions and other quantities
are then determined by simulating many nucleus-nucleus collisions.

%--- klaus_end:mc_approach_v1

%MLM --
\begin{figure}[tbp]
\begin{center}
\includegraphics[width=73mm]{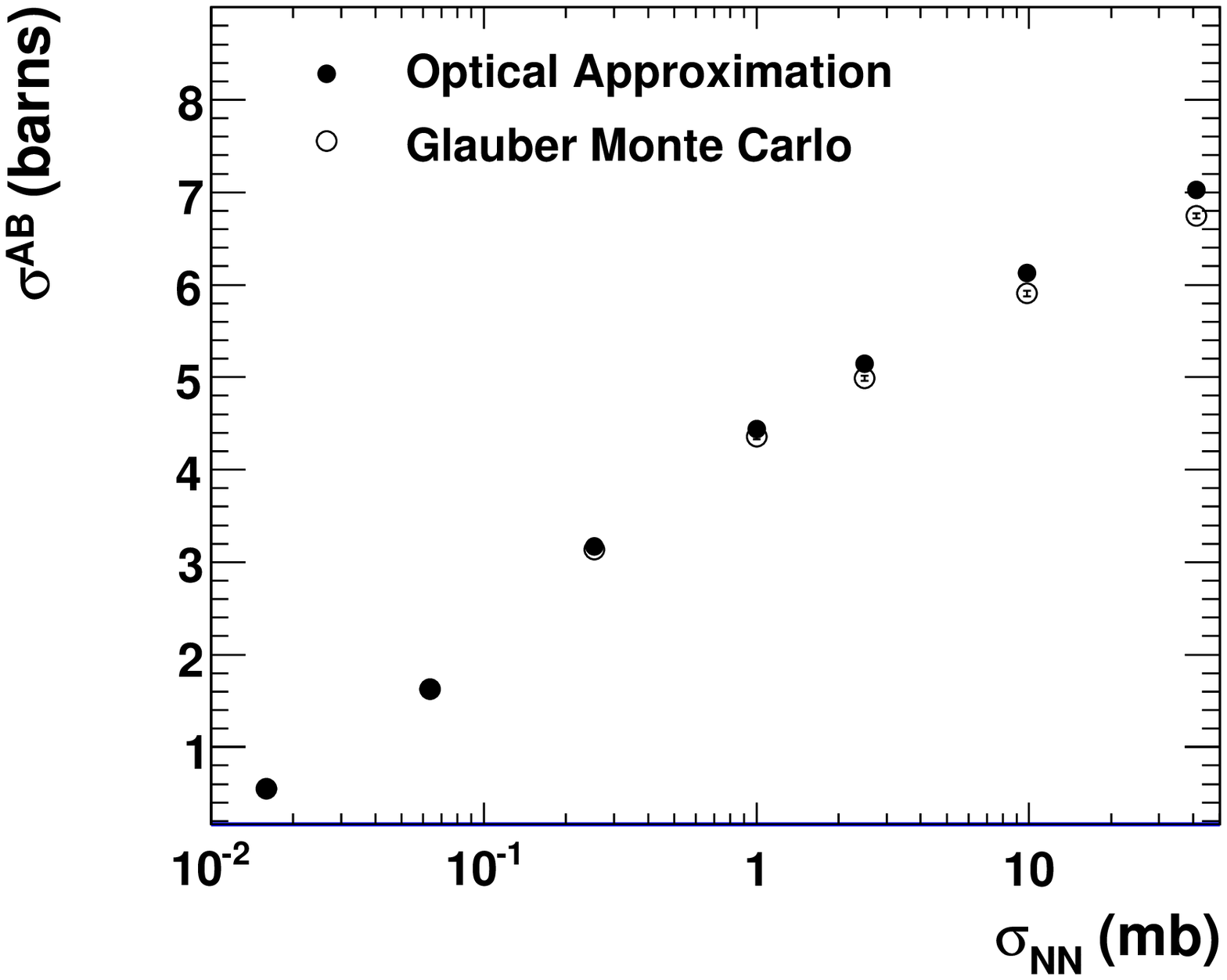}
\includegraphics[width=60mm]{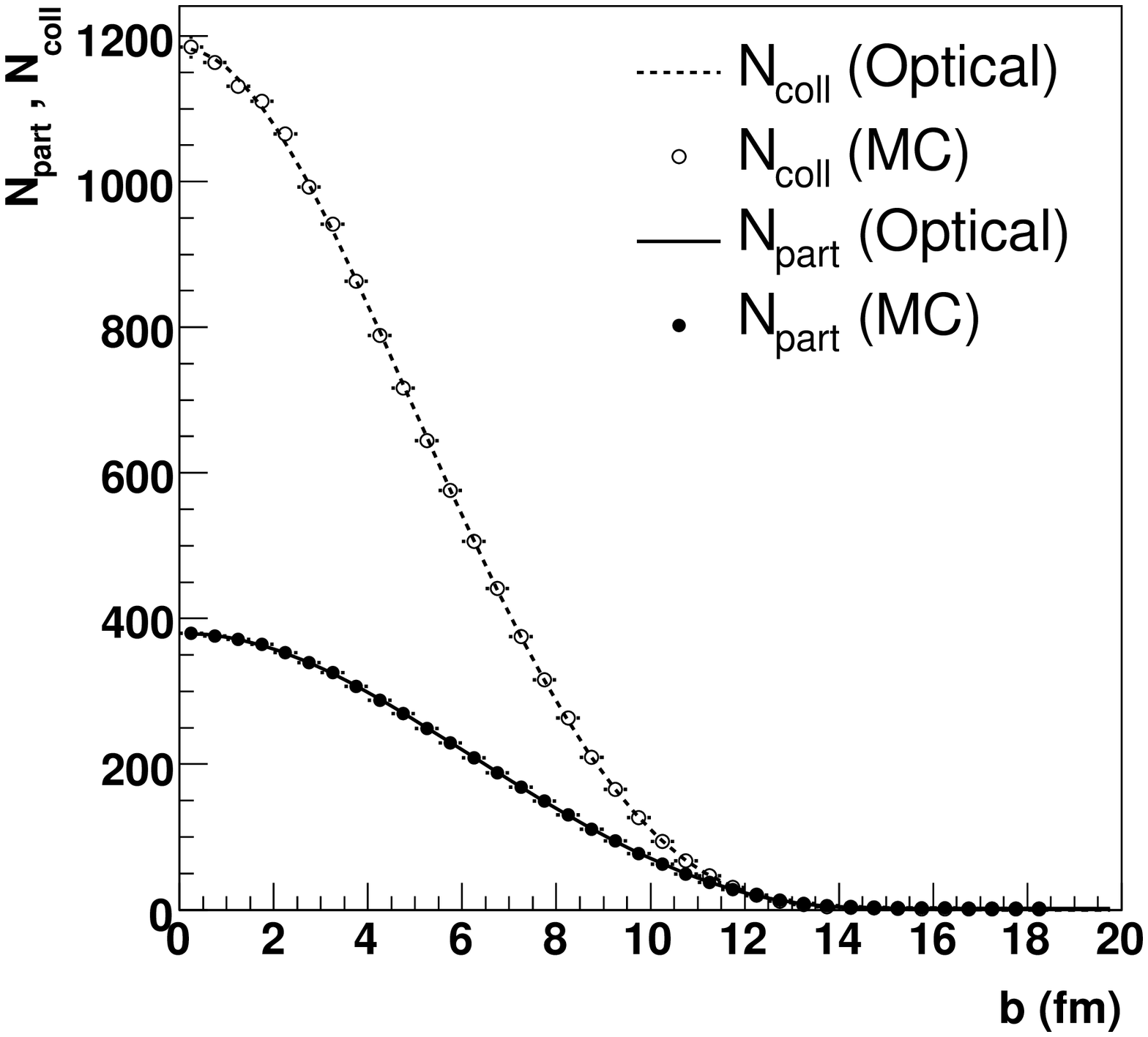}
\caption{
(left) Total cross section, calculated in the optical approximation
and with a Glauber Monte Carlo, both with identical nuclear
parameters, as a function of $\sigmann$, the nucleon-nucleon
inelastic cross section.
(right) $N_\mathrm{coll}$ and $N_\mathrm{part}$ as a 
function of impact parameter,
calculated in the optical approximation (lines) and with a
Glauber Monte Carlo (symbols).  The two are essentially identical out to
$b=2R_A$.
\label{fig:dilute_limit}
}
\end{center}
\end{figure}
\subsection{Differences between Optical and Monte Carlo Approaches}
It is not always remembered that the various integrals used to
calculated physical observables in the ``Glauber Model'' are
predicated on a particular approximation called the optical limit.
This limit assumes that scattering amplitudes can be described by an
eikonal approach, where the incoming nucleons see the target as a
smooth density.  This approach captures many features of the collision
process, but does not completely capture the physics of the total
cross section.  Thus, it tends to lead to distortions in the
estimation of $N_\mathrm{part}$ and $N_\mathrm{coll}$ compared to
similar estimations using the Glauber Monte Carlo approach.

This can be seen by simply looking at the relevant integrals.
The full (2A+2B+2)-dimensional integral to calculate the total cross section
is~\cite{Bialas:1976ed}:
\begin{eqnarray}
\sigma_{AB} = \int d^2 b \int d^2 s^A_1 \cdots d^2 s^A_A d^2 s^B_1 \cdots d^2 s^B_B \times \nonumber \\
\hat{T}_A({\bf s^A_1})\cdots \hat{T}_A({\bf s^A_A}) \hat{T}_B({\bf s^B_1})\cdots \hat{T}_B({\bf s^B_B}) \times\\
\left\{ 1 - \prod_{j=1}^{B} \prod_{i=1}^{A} [ 1 - \hat{\sigma} ({\bf b-s^A_i+s^B_j}) ] \right\} \nonumber
\end{eqnarray}
where $\hat{\sigma}({\bf s})$ is normalized such that 
$\int d^2 s \hat{\sigma}({\bf s}) = \sigmann$,
while the optical limit version of the same calculation is (cf. Equation 
\ref{eq:sigma_AB}):
\begin{equation}
\sigma_{AB} = \int d^2 b \left\{ 1-[1-\sigmann \hat{T}_{AB}(b)]^{AB}  \right\}
\end{equation}
These expressions are generally expected to be the same for large A (and B)
and/or when $\sigmann$ is sufficiently small~\cite{Bialas:1976ed}.  
The main difference between the two is that many terms in the full calculation
are missing in the optical limit calculation.  
These are the terms which describe local density fluctuations
event-by-event.   Thus, in the optical limit, each
nucleon in the projectile sees the oncoming target as a smooth density.

One can test this interpretation to first order by plotting the
total cross section as a function of $\sigmann$ for an optical
limit calculation as well as a GMC calculation with the same 
parameters, as shown in the left panel of Fig.~\ref{fig:dilute_limit}.
One sees that as the nucleon-nucleon cross section becomes more
point-like, the optical and GMC cross sections converge.  This
confirms the general suspicion that it is the ability of GMC
calculations to introduce ``shadowing'' corrections that reduces the
cross section relative to the optical calculations
\cite{Glauber:1955qq}

And yet, when calculating simple geometric quantities like 
$N_\mathrm{part}$ and $N_\mathrm{coll}$ as a function of impact parameter, 
there is little difference between the two calculations, as
shown in the right panel of Fig.~\ref{fig:dilute_limit}.  
The only substantial difference comes at the highest impact
parameters, something which will be discussed in Section
\ref{sect:totalgeom}.  Fluctuations are also sensitive to
this difference, but there is insufficient room in this review
to discuss more recent developments~\cite{Miller:2003kd}.

\subsection{Glauber Model Systematics}
%MLM --
\begin{figure}[btp]
\begin{center}
\includegraphics[width=120mm]{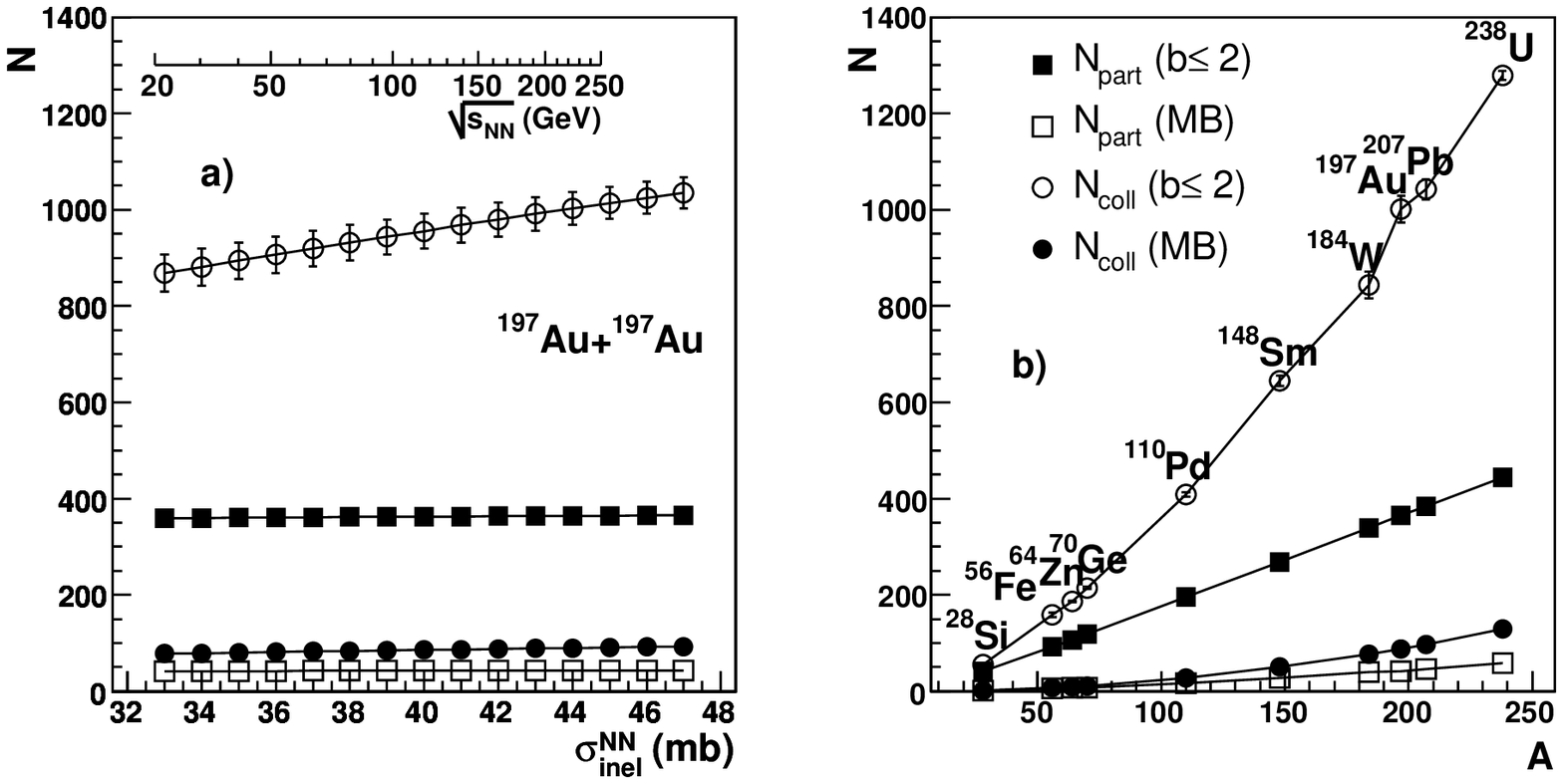}
\caption{
a) Dependence of $N_\mathrm{part}$ and $N_\mathrm{coll}$ on 
$\sigma^\mathrm{NN}_\mathrm{inel}$ for central ($b \leq 2~\mathrm{fm}$)
and minimum bias (MB) events for the $\mathrm{^{197}Au+^{197}Au}$ reaction.  
b) Dependence of these values on the system size, with
$\sigma^\mathrm{NN}_\mathrm{inel} = 42~\mathrm{mb}$. 
\label{fig:NSystematics.eps}
}
\end{center}
\end{figure}
%
%--- Steve_begin:systematics_v1
As discussed above,
the Glauber model depends on the nucleon-nucleon inelastic cross 
$\sigma^\mathrm{NN}_\mathrm{inel}$
section and the geometry of the interacting nuclei. In turn,
$\sigma^\mathrm{NN}_\mathrm{inel}$ depends on the energy of the reaction, as
shown in Fig.~\ref{fig:sigInelNN}, and the geometry depends on the
number of nucleons in the target and projectile.

Figure~\ref{fig:NSystematics.eps}a shows the effect of changing 
$\sigma ^\mathrm{NN}_\mathrm{inel} $ on the calculated values of the number of 
participant ($N_\mathrm{part}$)  and the number of collisions 
($N_\mathrm{coll}$) 
for $\mathrm{^{197}Au+^{197}Au}$ reaction over the range of 
$\sigma ^\mathrm{NN}_\mathrm{inel} $ values relevant
at RHIC. The secondary axis shows the values of $\sqrt{s_\mathrm{NN}}$
corresponding to the $\sigma ^\mathrm{NN}_\mathrm{inel} $ values.   
The values are shown for central events, 
with impact parameter $b < 2~\mathrm{fm}$, 
and for a minimum bias throw of events.   
The error bars,  which only extend beyond the symbol size for the 
$N_\mathrm{coll}$($b < 2~\mathrm{fm}$) results, are 
based on an assumed uncertainty of $\sigma _\mathrm{inel}^\mathrm{NN} $ 
at a given energy 
of 3~mb.  In general, one finds that the Glauber calculations show only
a weak energy dependence over the energy range covered by the RHIC 
accelerator.

Fig.~\ref{fig:NSystematics.eps}b shows dependence of $N_\mathrm{part}$ 
and $N_\mathrm{coll}$ on the system size for central and MB events, 
with values calculated for  identical particle 
collisions of the indicated systems at a fixed
value of $\sigma^\mathrm{NN}_\mathrm{inel} = 42~\mathrm{mb}$ (corresponding to
$\sqrt{s_\mathrm{NN}}=200~\mathrm{GeV}$). Since the Glauber Model is largely
dependent on the geometry of the colliding nuclei, some simple scalings can
be expected for $N_\mathrm{part}$ and $N_\mathrm{coll}$.

\begin{figure}[tbp]
\begin{center}
\includegraphics[width=120mm]{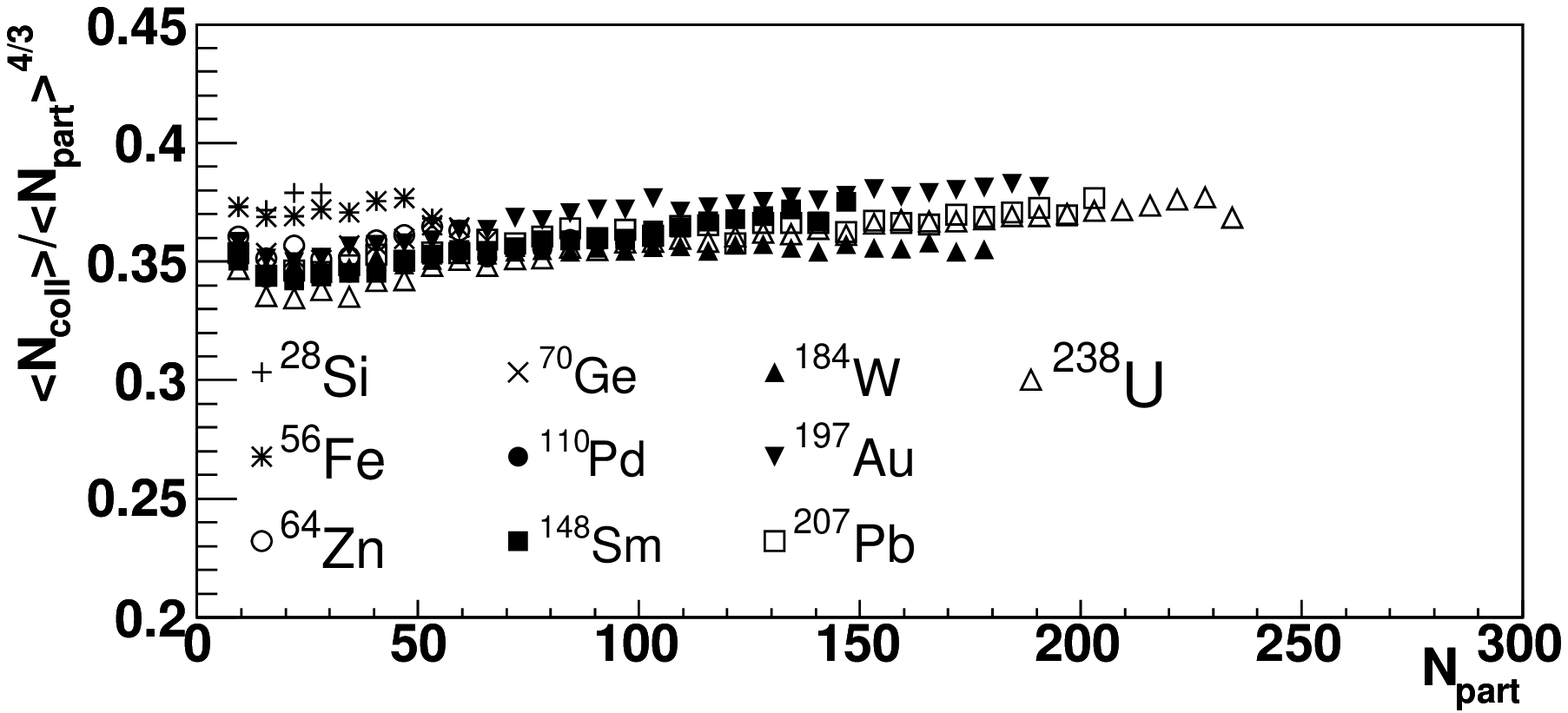}
\caption{
Geometric scaling behavior of $N_\mathrm{coll}$ as discussed in the text. 
The calculations are done with  $\sigma^\mathrm{NN}_\mathrm{inel} 
= 42~\mathrm{mb}$.  
\label{fig:NpartNcollScale.eps}
}
\end{center}
\end{figure}

$N_\mathrm{part}$ should scale with the volume of the interaction region.  
In Fig.~\ref{fig:NSystematics.eps} this is seen by the linear dependence of
$N_\mathrm{part}$ for central collisions on $A$, 
where the common volume $V$ of the largely overlapping nuclei in central collisions 
is proportional to $A$ for a saturated nuclear density. 
%$N_\mathrm{coll}$ also depends on the thickness of
%the interaction zone in the longitudinal direction, 
%as discussed in Sec. 3.4.2.  
%Still, for a given interaction volume, as reflected by
%$N_\mathrm{part}$, the ratio of $N_\mathrm{coll}$ 
%to $N_\mathrm{part}$ remains relatively 
%constant as a function of $A$.  
%
%The nucleon density in heavy nuclei is approximately constant so that
%the number of participants of a nucleus is proportional its
%interaction volume.  
In a collisions of two equal nuclei (A+A) the
average number of collisions per participant scales as the length $l_z
\propto N_\mathrm{part}^{1/3}$ of the interaction volume along the beam
direction so that the number of collisions roughly follows
\begin{equation}
N_\mathrm{coll} \propto N_\mathrm{part}^{4/3} \
\label{eq:ncollnpart}
\end{equation}
independent of the size of the nuclei.
This scaling relationship is demonstrated in 
Fig.~\ref{fig:NpartNcollScale.eps} 
where $N_\mathrm{coll} /N_\mathrm{part}^\mathrm{4/3}$
is plotted as a function of $N_\mathrm{part}$ 
for the systems shown in Fig.~\ref{fig:NSystematics.eps}b.
The result is confirmed experimentally in ref.~\cite{Alver:2005nb} 
where the $\mathrm{Cu+Cu}$ and $\mathrm{Au+Au}$ 
systems are compared. 
The geometric nature of the Glauber Model is evident.   

%\subsubsection{Charge distributions vs. Nucleon distributions}         
%The parameters used to characterize the nuclei going into Glauber
%calculations, both Monte Carlo and Optical, are typically taken
%from low energy electron scattering data.  By virtue of using
%electrons as the probe, these experiments measure the 
%electric charge distribution in great detail.  However, this
%may lead to some complications, since the nucleons in the
%colliding nuclei are extended objects with extended charge
%distributions.  
%Simple numerical calculations show that convoluting a charge
%distribution with radius $r_0$ with a spherically symmetric
%radial charge distribution generates an effective skin-depth
%of $a_{ch} \sim r_0/(2\sqrt{3})$.  This is added in 
%quadrature with $a$, the usual nuclear skin-depth parameter
%$a^2_{eff}= a^2 + a^2_{ch}$.  %

%While this has not been systematically considered in Glauber
%calculations, it should be a contribution to the systematic
%error.  Then again, the modification to the nominal distribution
%for Au involves replacing $a=0.535$ fm with $a_{eff} \sim 0.48$ fm.
%This lowers the total cross section calculated in the GMC
%approach by 3.7\%.  

%\begin{figure}[htbp]
%\begin{center}
%\caption{
%\label{fig:Ncollpart}
%}
%\end{center}
%\end{figure}

%\begin{figure}[htbp]
%\begin{center}
%\includegraphics[width=120mm]{sigma_tot_all_plus_optical.eps}
%\caption{ Total geometrical cross section from Glauber Monte Carlo
%  calculations (d+Au, Cu+Cu, and Au+Au at $\sqrt{s_\mathrm{NN}} =
%  200$~GeV).  The dashed line represents a optical limit calculation
%  for Au+Au.
%\label{fig:sigma_vs_b}
%}
%\end{center}
%\end{figure}

\clearpage

\newpage
\section{Relating the Glauber Model to Experimental Data}
\label{sect:experiment}

Unfortunately, neither $N_\mathrm{part}$ nor $N_\mathrm{coll}$ can be
directly measured in a RHIC experiment.  Mean values of such
quantities can be extracted for classes of ($N_\mathrm{evt}$) measured
events via a mapping procedure.  Typically a measured distribution
(e.g., $\mathrm{d}N_\mathrm{evt}/\mathrm{d}N_\mathrm{ch}$) is mapped to
the corresponding distribution obtained from phenomenological Glauber
calculations.  This is done by defining ``centrality classes'' in both
the measured and calculated distributions and then connecting the mean
values from the same centrality class in the two distributions.  The
specifics of this mapping procedure differ both between experiments as
well as between collision systems within a given experiment.  Herein
we briefly summarize the principles and various implementations of
centrality definition.

\subsection{\label{sec:basicMeth}Methodology}
\begin{figure}[htbp]
\begin{center}
\includegraphics[width=80mm]{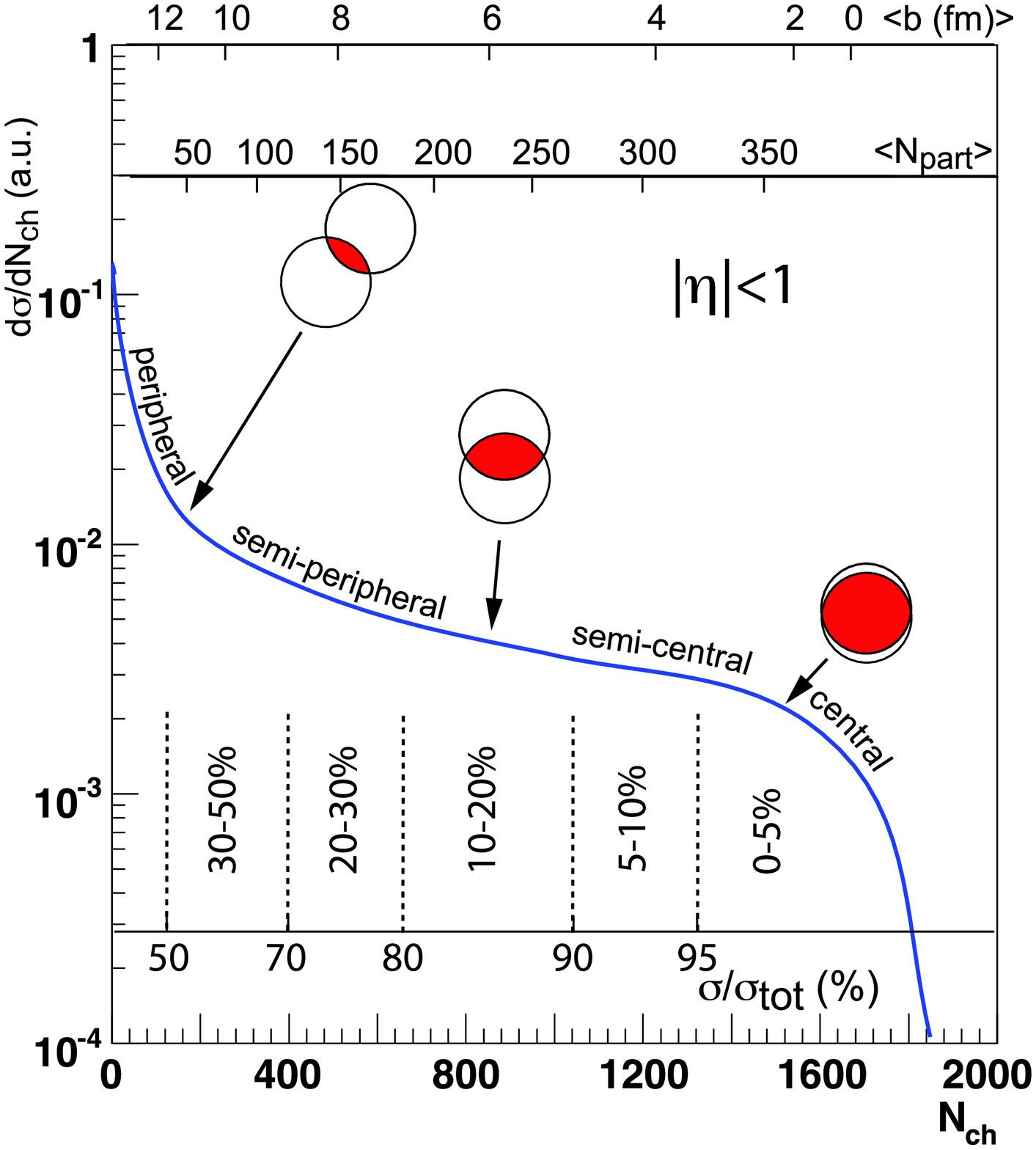}
\caption{
A cartoon example of the correlation of the final state observable $N_\mathrm{ch}$ with Glauber calculated quantities ($b$, $N_\mathrm{part}$).  The plotted distribution and various values are illustrative and not actual measurements (T. Ullrich, private communication).  
\label{fig:cocktail}
}
\end{center}
\end{figure}
The basic assumption underlying centrality classes is that the impact
parameter $b$ is monotonically related to particle multiplicity, both
at mid and forward rapidity.  For large $b$ events (``peripheral'') we
expect low multiplicity at mid-rapidity, and a large number of
spectator nucleons at beam rapidity, whereas for small $b$ events
(``central'') we expect large multiplicity at mid-rapidity and a small
number of spectator nucleons at beam rapidity
(Figure~\ref{fig:cocktail}).  In the simplest case, one measures the
per-event charged particle multiplicity
($\mathrm{d}N_\mathrm{evt}/\mathrm{d}N_\mathrm{ch}$) for an ensemble
of events.  Once the total integral of the distribution is known,
centrality classes are defined by binning the distribution based upon
the fraction of the total integral.  The dashed vertical lines in
Figure~\ref{fig:cocktail} represent a typical binning.  The same
procedure is then applied to a \textit{calculated} distribution, often
derived from a large number of Monte Carlo trials.  For each
centrality class, the mean value of Glauber quantities (e.g., $\langle N_{part} \rangle$)
for the Monte Carlo events within the bin (e.g., 5-10\%) is
calculated.  Potential complications to this straightforward procedure
arise from various sources: event selection, uncertainty in the total
measured cross section, fluctuations in both the measured and
calculated distributions, and finite kinematic acceptance.

\subsubsection{Event Selection}

All four RHIC experiments share a common detector to select minimum
bias (MB) heavy ion events.  The Zero Degree Calorimeters (ZDCs) are
small acceptance hadronic calorimeters with angular coverage of
$\theta \le 2$~mrad with respect to the beam axis \cite{Adler:2000bd}.
Situated behind the charged particle steering DX magnets of RHIC, the
ZDCs are primarily sensitive to neutral spectators.  For Au+Au
collisions at $\sqrt{s_\mathrm{NN}} = 130$~GeV and above, the ZDCs are
$\sim 100$\% efficient for inelastic collisions, thus providing an
excellent MB trigger.  The RHIC experiments often apply an online
timing cut to select events within a given primary vertex interval
($\sim |z_\mathrm{vertex}|<$30 cm).  Further coincidence with fast
detectors near mid-rapidity are often also used to suppress background
events such as beam-gas collisions.  Experiment specific event
selection is described in detail in section \ref{sec:centDetails}.
%MLM
 \begin{figure}[tbp]
\begin{center}
\includegraphics[width=120mm]{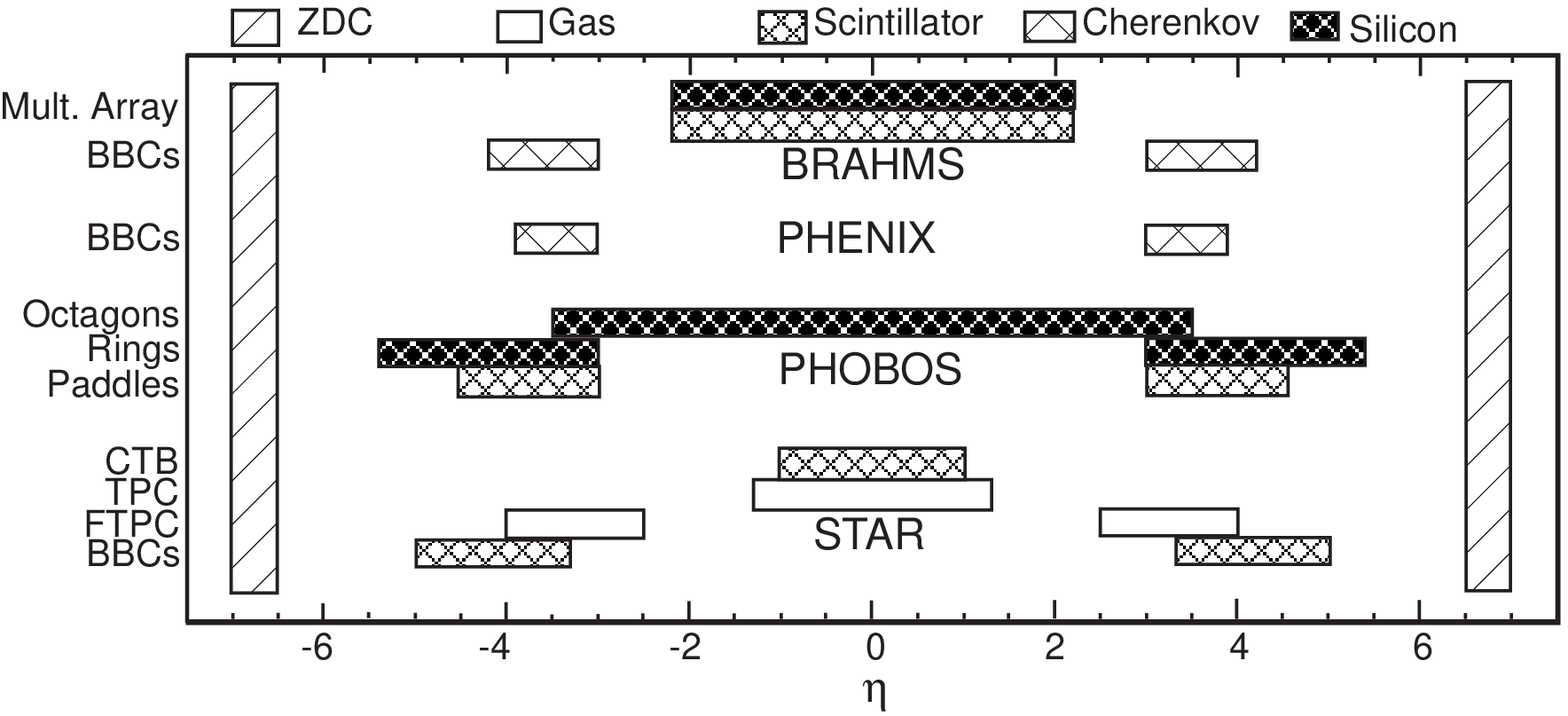}
\caption{
Pseudorapidity coverage of the centrality selection detectors of the 4 RHIC experiments.  Top to bottom: BRAHMS, PHENIX, PHOBOS, STAR.  
\label{fig:coverage}
}
\end{center}
\end{figure}
Figure~\ref{fig:coverage} displays the pseudorapidity coverage of the
suite of subsystems used to define centrality (both offline and at the
trigger level) \cite{Adamczyk:2003sq, Allen:2003zt, Back:2003sr,
  Braem:2002ae}.  With the exception of the STAR TPC and FTPCs, all
other subsystems are intrinsically fast and available for event
triggering.
\subsubsection{Centrality Observables}
In minimum bias p+p and p+$\bar{\mathrm{p}}$ collisions at high
energy, the charged particle multiplicity $\mathrm{d}N_\mathrm{evt}
/\mathrm{d}N_\mathrm{ch}$ has been measured over a wide range of
rapidity and is well described by a negative binomial distribution
\cite{Ansorge:1988kn}.  However, the multiplicity is also known to
scale with the hardness ($q^2$) of the collision -- the multiplicity
for hard jet events is significantly higher than that of MB
collisions.  In heavy ion collisions, we manipulate the fact that the
majority of the initial state nucleon-nucleon collisions will be
analogous to minimum bias p+p collisions, with a small perturbation
from much rarer hard interactions.  The final integrated multiplicity
of heavy ion events is then roughly described as a superposition of
many negative binomial distributions, which quickly approaches the
Gaussian limit.
% It critical to note that the study of high $p_T$ events from
% peripheral heavy ion collisions is a dicey business.

$N_\mathrm{ch}$ can be measured offline by counting charged tracks
(e.g, STAR TPC) or estimated online from the total energy deposited in
a detector divided by the typical energy deposition per charged
particle (e.g., PHOBOS Paddles).  
 \begin{figure}[tbp]
\begin{center}
\includegraphics[width=120mm]{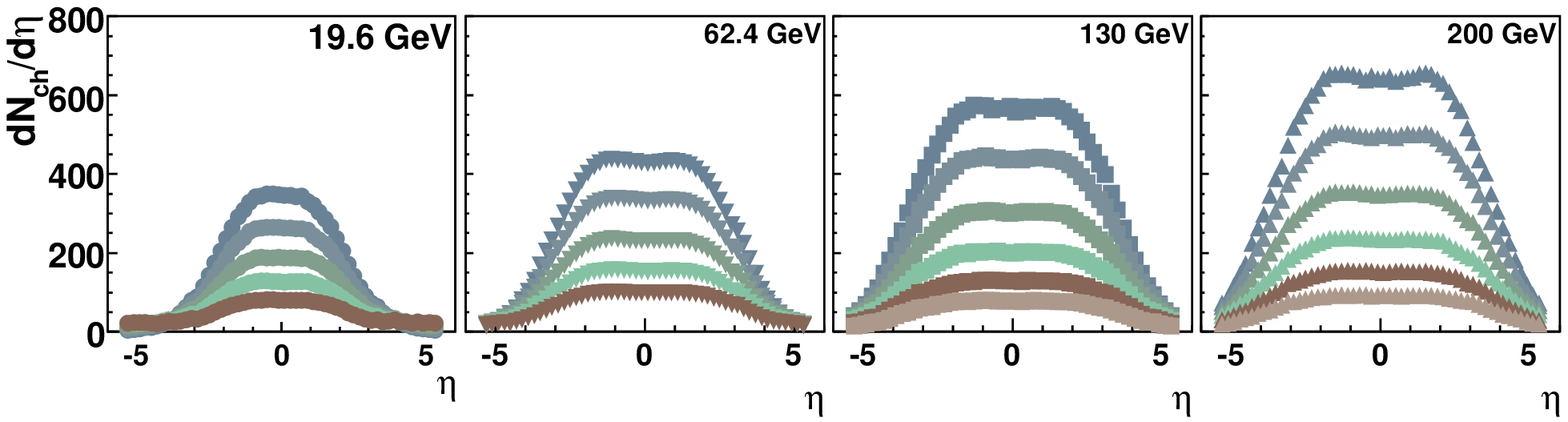}
\caption{
Charged particle multiplicity (PHOBOS) in Au+Au collisions 
at various center-of-mass energies for $|\eta|<5$
\cite{Back:2002wb, Back:2005hs}.  The different colors 
represent different centrality selections.
\label{fig:phobosNcharge}
}
\end{center}
\end{figure}
As shown in
Figure~\ref{fig:coverage}, there is a wide variation in the acceptance
of various centrality detectors at RHIC.  PHOBOS, with the largest
acceptance in $\eta$, is well suited to measure
$\frac{\mathrm{d}^2N_\mathrm{evt}}{\mathrm{d}N_\mathrm{ch}
  \mathrm{d}\eta}$ (Figure~\ref{fig:phobosNcharge}).  These data illustrate two key features of particle production in nucleus-nucleus
collisions.  At a fixed beam energy, there is no dramatic change in
shape as the centrality changes.  However reducing the beam energy
{\it does} change the shape substantially, since the maximum rapidity
varies as $\ln(\sqrt{s_\mathrm{NN}})$.  Thus, the same trigger
detector may have a very different overall efficiency at different
beam energies.

$N_\mathrm{ch}$ can be simulated via various methods, but all require
the coupling of a Glauber calculation to a model of charged particle
production, either dynamic (e.g. HIJING \cite{hijing}) or static
(randomly sampled from a Gaussian, Poisson, or negative binomial).
Most follow the general prescription that the multiplicity scales
approximately with $N_\mathrm{part}$.  For an optical Glauber
calculation, simulated multiplicity ($N_\mathrm{ch}^\mathrm{sim}$) can
be calculated semi-analytically assuming that each participant
contributes a given value of $N_\mathrm{ch}$ which is typically drawn
from one of the aforementioned static probability distributions
\cite{Kharzeev:2000ph}.  The same can be done for a Monte Carlo
Glauber simulation with the added advantage that the detector response
to such ``events" can be simulated, thus enabling an apples-to-apples
comparison of simulated and measured $N_\mathrm{ch}$ distributions.
Various dynamical models of heavy ion collisions exist and can also be
coupled to detector simulations.  In all cases, the exact value of
$N_\mathrm{part}$, $N_\mathrm{coll}$, $b$, and
$N_\mathrm{ch}^\mathrm{sim}$ are known for each event.

\subsubsection{\label{sec:divide}Dividing by Percentile of Total Inelastic Cross Section}
With a measured and simulated $\mathrm{d}N_\mathrm{evt}/
\mathrm{d}N_\mathrm{ch}$ distribution in hand, one can then perform
the mapping procedure to extract mean values.  Suppose that the
measured and simulated distributions are both one dimensional
histograms.  For each histogram, the total integral is calculated and
centrality classes are defined in terms of fraction of the total
integral.  Typically the integration is performed from large values of
$N_\mathrm{ch}$ to small.  For example, the 10-20\% \textit{most
  central} class is defined by boundaries $n_{10}$ and $n_{20}$ which
satisfy
\begin{equation}
  \frac{ \int_{\infty}^{n_{10}} \frac{\mathrm{d}N_\mathrm{evt}}{\mathrm{d}N_\mathrm{ch}} 
    \, \mathrm{d}N_\mathrm{ch} } {\int_{\infty}^0 \frac{\mathrm{d}N_\mathrm{evt}}
    {\mathrm{d}N_\mathrm{ch}} \, \mathrm{d}N_\mathrm{ch} }=0.1
\textrm{    and    }
\frac{ \int_{\infty}^{n_{20}} \frac{\mathrm{d}N_\mathrm{evt}}{\mathrm{d}N_\mathrm{ch}} 
  \, \mathrm{d}N_\mathrm{ch} } {\int_{\infty}^0 \frac{\mathrm{d}N_\mathrm{evt}}{\mathrm{d}N_\mathrm{ch}} 
  \, \mathrm{d}N_\mathrm{ch} }=0.2 \,.
\label{eq:centralityClass}
\end{equation}
See, for example, Figure~\ref{fig:cocktail}.  The same procedure is
performed on both the measured and simulated distribution. We note
explicitly that $n_{i}^\mathrm{measured}$ need not equal
$n_{i}^\mathrm{simulated}$.  This non-trivial fact implies that the
mapping procedure is robust to an overall scaling of the simulated
$N_\mathrm{ch}$ distribution compared to the measured distribution.
%MLM --
\begin{figure}[tbp]
\begin{center}
\includegraphics[width=100mm]{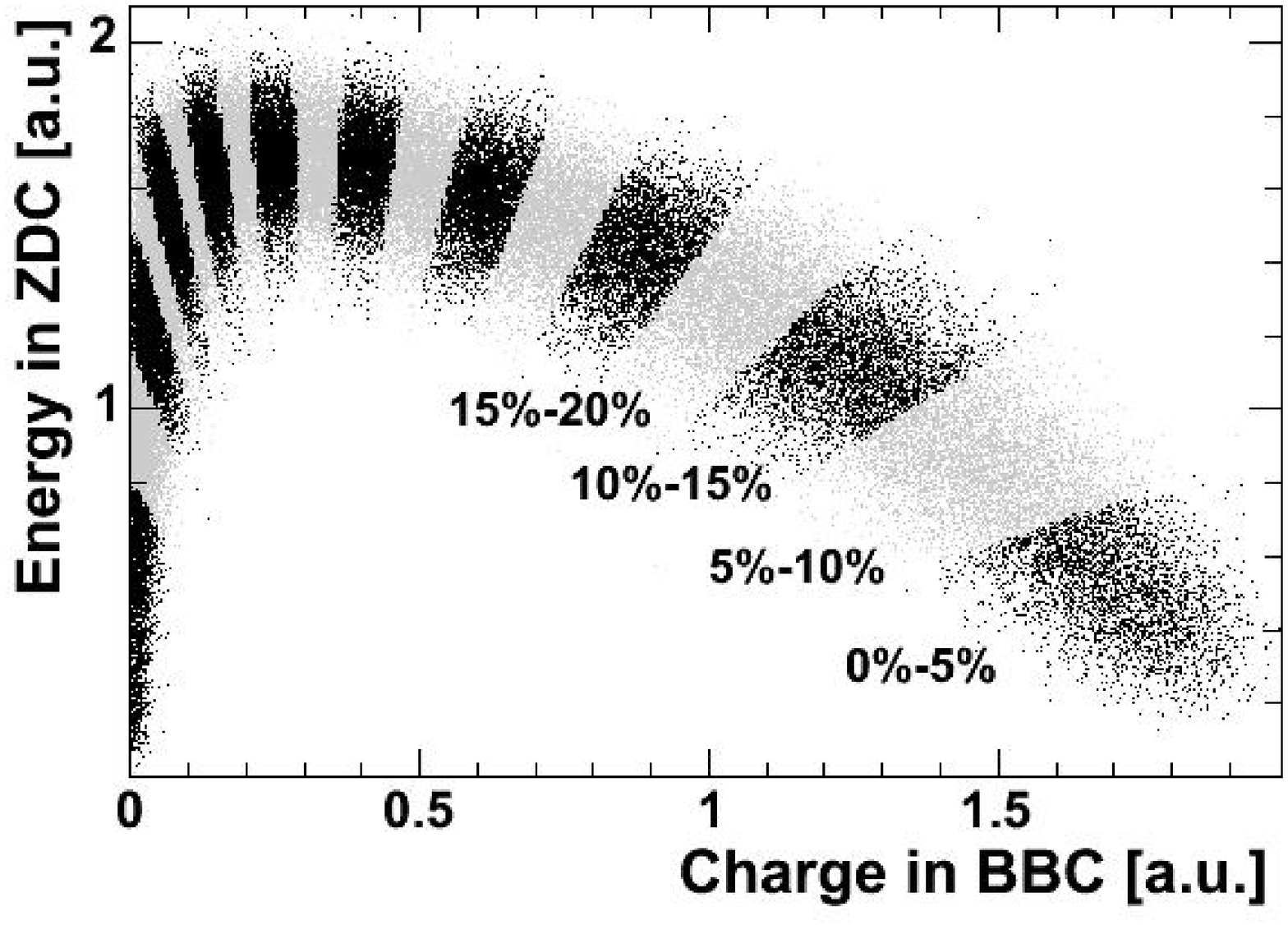}
\caption{
Spectator energy deposition in the ZDCs as a function of charged particle 
multiplicity in the BBCs (PHENIX).
\label{fig:phxBoomerang}
}
\end{center}
\end{figure}
One can extend the centrality classification beyond one dimension by
studying the correlation of beam rapidity spectator multiplicity with
mid-rapidity particle production (Figure~\ref{fig:phxBoomerang}).  Although the distribution is somewhat asymmetric, the naive
expectations of Section \ref{sec:basicMeth} are clearly upheld and the
mapping procedure proceeds as in the 1-d case.

Once a centrality class is defined in simulation, the mean values of quantities such as 
$N_\mathrm{part}$ can be calculated for events that fall in
that centrality bin.  Systematic uncertainty in the total measured
cross section propagates into a leading systematic uncertainty on the
Glauber quantities extracted via the mapping
process.  This uncertainty can be directly propagated by varying the
value of the denominator in Equation \ref{eq:centralityClass}
accordingly and recalculating $\langle N_\mathrm{part} \rangle$, etc.
For example, for Au+Au collisions at $\sqrt{s_\mathrm{NN}}=200$ GeV in
the 10-20\% (60-80\%) most central bin, the STAR collaboration quotes
values of $\langle N_\mathrm{part} \rangle \approx$ 234 (21) with an
uncertainty of $\sim$ 6 (5) from the 5\% uncertainty on the total
cross section alone \cite{Adler:2002xw}.  Clearly the uncertainty in
total cross section becomes increasingly important as one approaches
the most peripheral collisions.

\subsection{\label{sec:centDetails}Experimental Details}
\subsubsection{BRAHMS}
The BRAHMS experiment
uses the charged particle multiplicity observed in a pseudorapidity
range of $ - 2.2 \le \eta \le 2.2$ to determine reaction centrality
\cite{Bearden:2001xw,Bearden:2001qq,Arsene:2004cn}.  The
multiplicities are measured in a ``multiplicity array'' consisting of
an inner barrel of Si strip detectors (SiMA) and an outer barrel of
plastic scintillator ``tile'' detectors (TMA)
\cite{Adamczyk:2003sq,Lee:2004su} for collisions within 36~cm of the 
nominal vertex location.  Both arrays cover the same
pseudorapidity range for collisions at the nominal vertex.   
``Beam-Beam''-counter arrays (BBC)located on either side of the nominal
interaction point at a distance of 2.2 m of the nominal vertex 
extend the  pseudorapidity coverage for measuring
charge-particle pseudorapidity densities.  These arrays consist of Cherenkov
UV-transmitting plastic radiators coupled to photomultiplier tubes.
%MLM --
\begin{figure}[tbp]
\begin{center}
\includegraphics[width=80mm]{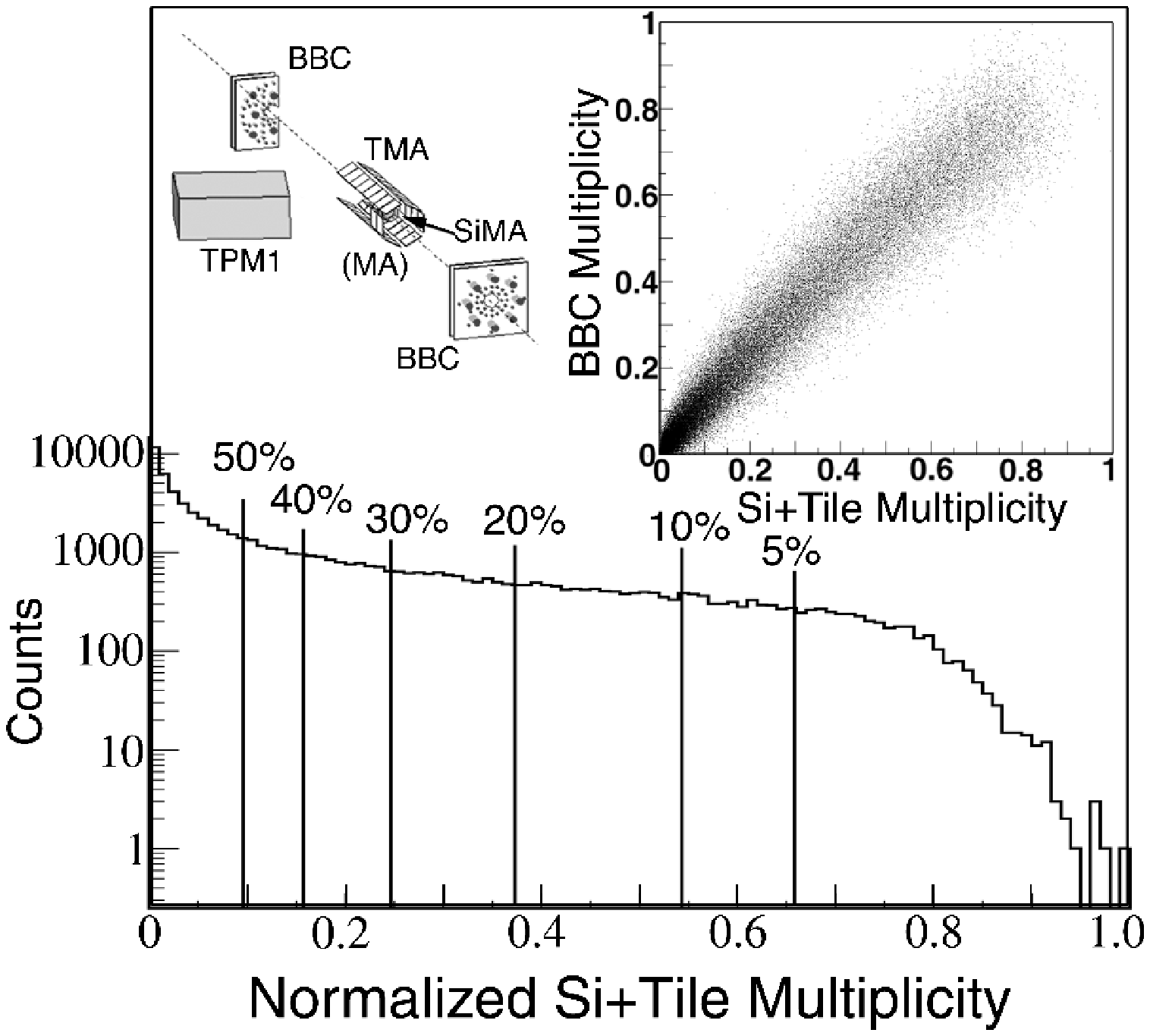}
\caption{
Normalized multiplicity distribution for the $^{197}\rm{Au}+^{197}\rm{Au}$ 
reaction at $\sqrt{s_\mathrm{NN}} = 130~\rm{GeV}$
measured in the BRAHMS multiplicity array\cite{Bearden:2001xw}.  The insert shows the correlation pattern 
of multiplicities measured in the Beam-Beam Counter arrays and the multiplicity array.  
The vertical
lines indicate multiplicity values associated with the indicated centrality cuts.
\label{fig:BrahmsMult}
}
\end{center}
\end{figure}
Figure~\ref{fig:BrahmsMult} shows the normalized multiplicity
distribution measured in the multiplicity array for the
$^{197}\rm{Au}+^{197}\rm{Au}$ reaction at $\sqrt{s_\mathrm{NN}} =
130~\rm{GeV}$.   The insert shows a correlation plot of the
multiplicity measured in the Beam-Beam counter array and that in the
multiplicity array. The vertical lines indicate the multiplicities
corresponding to the indicated centrality values.

The BRAHMS reference multiplicity distribution requires coincident
signals in the experiment's two ZDC detectors, an interaction vertex
located within 30 cm of the nominal vertex location, and that there be
at least four ``hits'' in the TMA. This additional requirement largely
removes background contributions from beam-residual gas interactions
and from very peripheral collisions involving only electromagnetic
processes.  The collision vertex can be determined by either the BBC-arrays,
the ZDC counters, or a time-projection chamber that is part of a mid-rapidity
spectrometer arm (TPM1 in Fig.~\ref{fig:BrahmsMult}).

A simulation of the experimental response based on realistic GEANT3
simulations \cite{geant3} and using the HIJING Monte Carlo event
generator \cite{hijing} for input was used to estimate the fraction of
the inelastic scattering yield that was missed in the experiment's
minimum-bias event selection.  Multiplicity spectra using the
simulated events are compared to the experimental spectra. The shapes
of the spectra are found to agree very well for an extended range of
multiplicities in the TMA array above the threshold multiplicity set
for the event selection. The simulated events are then used to
extrapolate the experimental spectrum below the threshold.  Using this
procedure, it is estimated that the minimum-bias event selection
criteria selects $(93 \pm 3){\rm{\% }}$ of the total nuclear cross
section for Au+Au collisions at $\sqrt {s_\mathrm{NN} } = 200{\kern
  1pt} \;{\rm{GeV}}$, down to $(87 \pm 7){\rm{\% }}$ for Cu+Cu
collisions at $\sqrt {s_\mathrm{NN} } = 62.4{\kern 1pt} \;{\rm{GeV}}$.
%--- steve_end

\subsubsection{PHENIX}

% --- klaus_begin:sigma_tot_fraction_v1
We consider two examples: Au+Au at $\sqrt{s_\mathrm{NN}} = 200$~GeV
and Cu+Cu at $\sqrt{s_\mathrm{NN}} = 22.4$~GeV. The minimum bias
trigger condition for Au+Au collisions at $\sqrt{s_\mathrm{NN}} =
200$~GeV was based on Beam-Beam-Counters (BBCs) \cite{Allen:2003zt}.
The two BBCs ($3.1 \le |\eta| \le 3.9$) each consist of 64
photomultipliers which detect Cherenkov light produced by charged
particles traversing quartz radiators. On the hardware level a minimum
bias event was required to have at least 2 photomultiplier hits in
each BBC. Some analyses only used events with an additional hardware
coincidence of the two ZDCs. Moreover, the interaction vertex along
the beam axis ($z$ axis) reconstructed based on the arrival time
difference in the two BBCs was required to lie within $\pm 30$~cm of
the nominal vertex.

The efficiency of accepting inelastic Au+Au collisions under the
condition of having at least two photomultiplier hits in each BBC
($N_\mathrm{PMT}^\mathrm{BBC} \ge 2$) was determined with the aid of
HIJING Monte Carlo events \cite{hijing} and a detailed simulation of
the BBC response \cite{geant3}.  With an offline vertex cut of $\pm
30$~cm these simulations yield an efficiency of $(92.3 \pm 2)\,\%$ for
Au+Au at $\sqrt{s_\mathrm{NN}} = 200$~GeV \cite{Adler:2004zn}.  The
systematic uncertainties reflect uncertainties of (a)
$\mathrm{d}N_\mathrm{ch}/\mathrm{d}y$ in HIJING, (b) the shape of the
$z$-vertex distribution, and (c) the stability of the photomultiplier
gains.

The additional requirement of a ZDC coincidence removes remaining
background from beam-gas interactions, but possibly also leads to a
small inefficiency for peripheral collisions. The efficiency of
accepting real Au+Au collisions with $N_\mathrm{PMT}^\mathrm{BBC} \ge
2$ under the condition of a coincidence of the ZDCs was estimated to
be $99^{+1.0}_{-1.5}$~\%. Combining the efficiencies of the BBC and
ZDC requirement as well as the offline vertex cut PHENIX finds that
its sample of minimum bias events in Au+Au at $\sqrt{s_\mathrm{NN}} =
200$~GeV corresponds to $91.4 ^{+2.5}_{-3.0}\,\%$ of the total
inelastic cross section. Centrality classes were then defined by cuts
on the two-dimensional distribution of the ZDC energy as a function of
the BBC signal as shown in Figure~\ref{fig:phxBoomerang}.
% --- klaus_end:sigma_tot_fraction_v1

% --- klaus_begin:nbd
As a second example we consider the centrality selection in Cu+Cu
collisions at $\sqrt{s_\mathrm{NN}} = 22.4$~GeV. At this energy the
beam rapidity $y_\mathrm{beam} \approx 3.2$ lies within the
pseudorapidity range of the BBCs. The BBCs were still used as minimum
bias trigger detectors ($N_\mathrm{PMT}^\mathrm{BBC} \ge 1$ in both
BBCs). However, a monotonic relation between the BBC signal and the
impact parameter was no longer obvious.  Thus, the hit multiplicity
($N_\mathrm{PC1}$) measured with a Pad Chamber detector
\cite{Aronson:2003zn} at mid-rapidity ($|\eta| \le 0.35$) was used as
centrality variable.

%MLM --
\begin{figure}[tbp]
\begin{center}
\includegraphics[width=100mm]{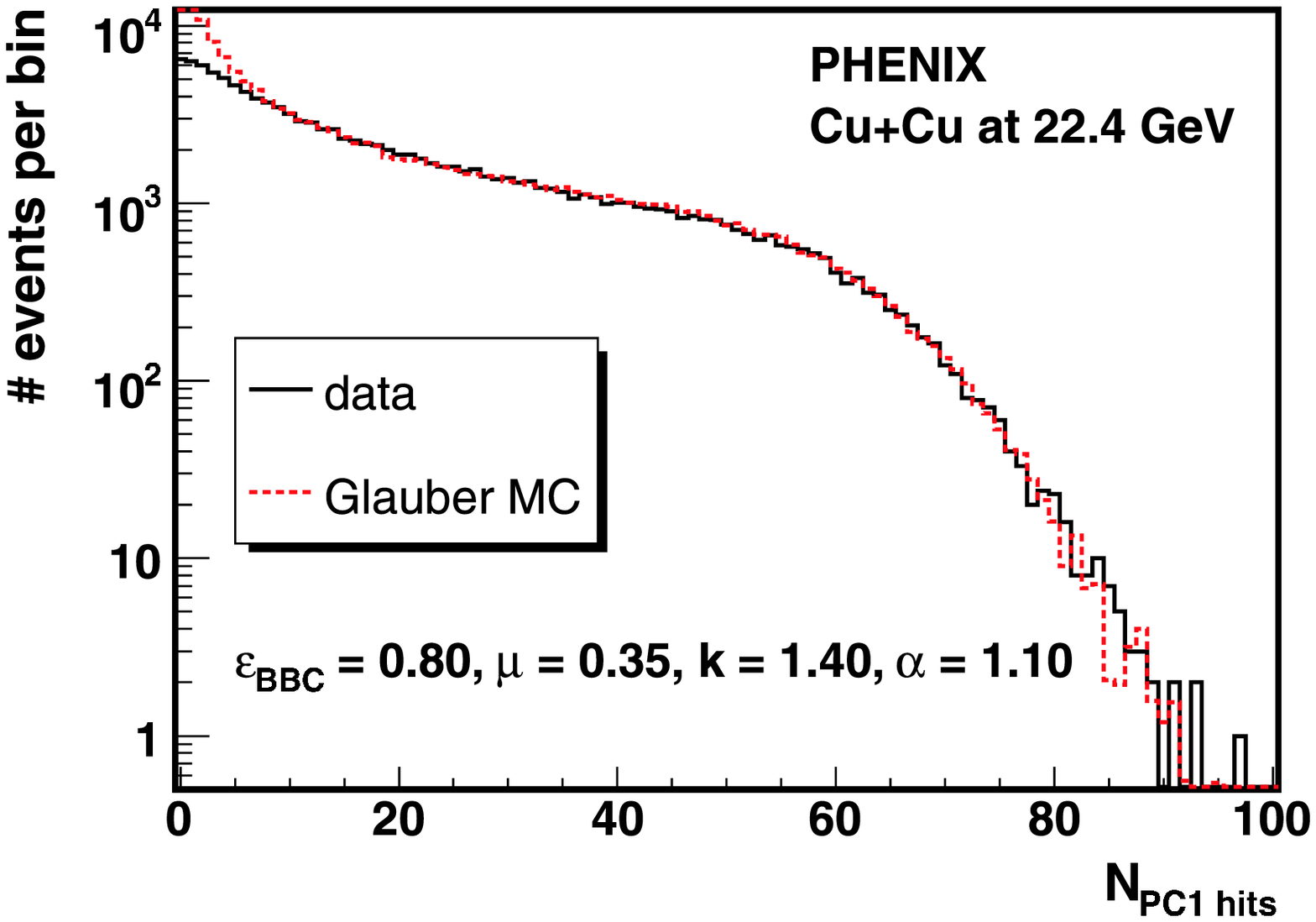}
\caption{
Measured and simulated distribution of the Pad Chamber 1 hit 
multiplicity used as centrality variable in Cu+Cu collisions 
at $\sqrt{s_\mathrm{NN}} = 22.4$~GeV (PHENIX).
\label{fig:npc1_cucu_22gev}
}
\end{center}
\end{figure}
The PC1 multiplicity distribution was simulated based on a convolution
of the $N_\mathrm{part}$ distribution from Glauber MC and a negative
binomial distribution (NBD). A non-linear scaling of the average
particle multiplicity with $N_\mathrm{part}$ was allowed: it was
assumed that the number of independently decaying precursor particles
('ancestors', $N_\mathrm{ancestor}$) is given by $N_\mathrm{ancestor}
= N_\mathrm{part}^\alpha$. The number of measured PC1 hits per
precursor particle was assumed follow a NBD:
\begin{equation}
  P_{\mu,k}(n) = \frac{\Gamma(n+k)}{\Gamma(n+1) \, \Gamma(k)} \cdot
  \frac{(\mu/k)^n}{(\mu/k + 1)^{n+k}} \, .
\label{eq:nbd}
\end{equation}
In a Glauber MC event the NBD was sampled $N_\mathrm{ancestor}$ times
to obtain the simulated PC1 multiplicity for this event.  The PC1
multiplicity distribution was simulated for a grid of values for
$\mu$, $k$ , and $\alpha$ in order to find optimal values.
Figure~\ref{fig:npc1_cucu_22gev} shows the measured and simulated PC1
distribution along with the best estimate of the BBC trigger
efficiency ($\varepsilon_\mathrm{BBC} \approx 0.8$) corresponding to
the difference at small $N_\mathrm{PC1}$ (see \cite{Adler:2004zn} for
a similar study in Au+Au collisions at $\sqrt{s_\mathrm{NN}} =
19.6$~GeV).  Given the good agreement between the measured and
simulated distribution in Figure~\ref{fig:npc1_cucu_22gev} centrality
classes for Cu+Cu collisions at $\sqrt{s_\mathrm{NN}} = 22.4$~GeV were
defined by identical cuts on the measured and simulated
$N_\mathrm{PC1}$.
% --- klaus_end:nbd

\subsubsection{PHOBOS}

As discussed above, PHOBOS measures centrality with 2 sets of 16
scintillator paddle counters covering $3<|\eta|<4.5$
\cite{Back:2003sr}.  Good events are defined by having less than 4 ns
time difference between the first hit impinging on each paddle counter
(limiting the vertex range) and either a coincidence between the
PHOBOS ZDCs or a high summed energy signal in the paddles (to avoid
the slight inefficiency in the ZDCs at small impact parameter at low
energies).

%MLM --
\begin{figure}[tbp]
\begin{center}
\includegraphics[width=100mm]{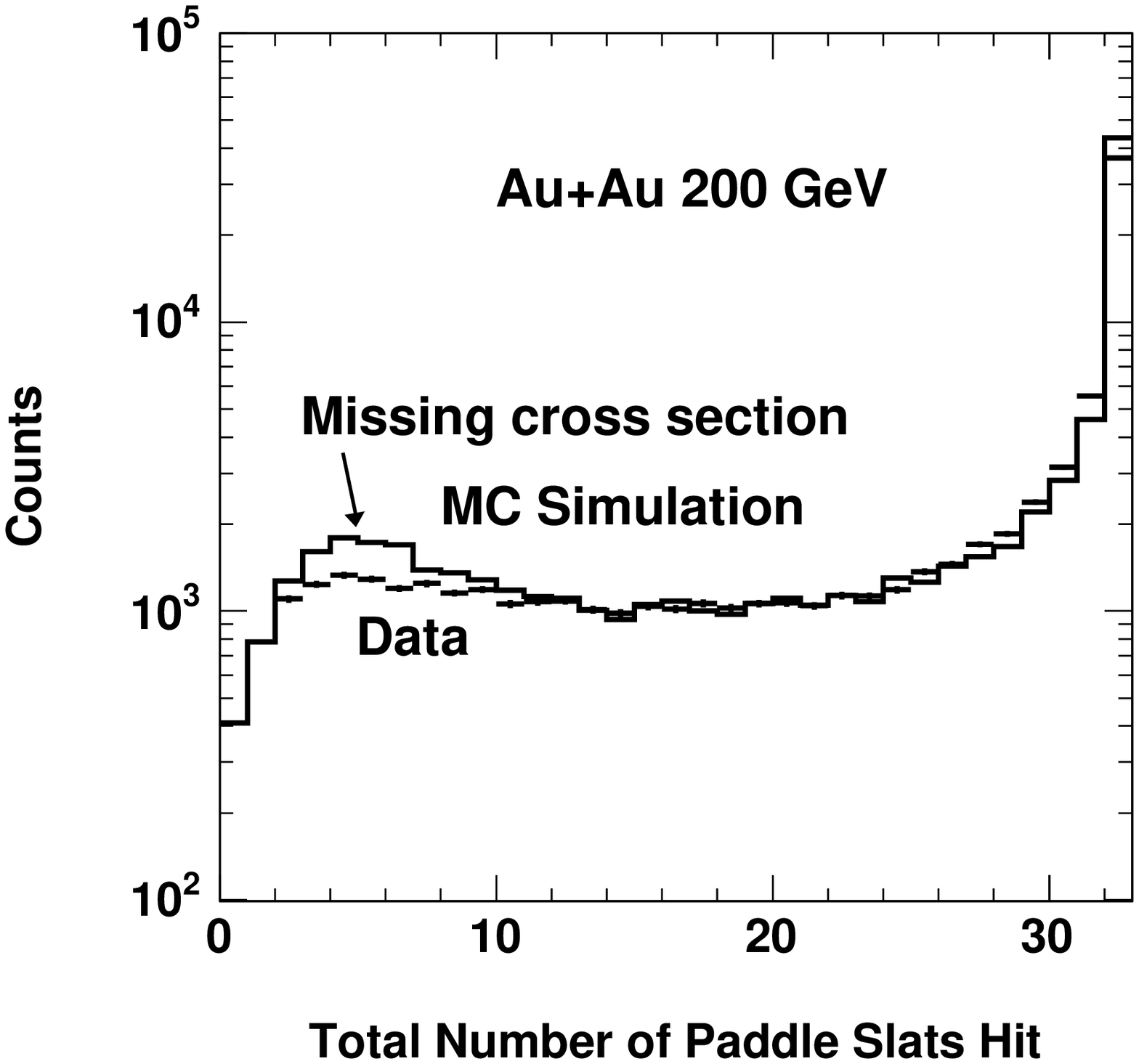}
\caption{
\label{fig:WP38_TotPaddleHit_Data_MC}
PHOBOS data showing the number of paddle slats hit by
charged particles, out of 32 total, in triggered events.
The data is compared with a full Monte Carlo simulation
of unbiased HIJING events.  The data is normalized around
$N(slats)\sim 16$ in order to compare data and MC, to 
estimate the fraction of events lost to the trigger
conditions.
}
\end{center}
\end{figure}
PHOBOS estimates the observed fraction of the total cross section by
measuring the distribution of the total number of paddle slats, as
shown in Figure~\ref{fig:WP38_TotPaddleHit_Data_MC}.  
Most of the variation in this quantity essentially measures the very
low-multiplicity part of the multiplicity distribution, since the bulk
of the events have sufficient multiplicity to fire all of the paddles
simultaneously.  The inefficiency is determined by matching the
``plateau'' structure in the data to that in HIJING, and measuring how
many of the low multiplicity events are missed in the data relative to
the MC calculation.  This accounts for a variety of instrumental
effects in an aggregate way, and the difference between the estimated
value and 100\% sets the scale for the systematic uncertainty.

The total detection efficiency for 200 GeV Au+Au collisions is found
to be 97\%, and 88\% when requiring more than 2 slats hit on each set
of 16 paddle counters.  This last requirement dramatically reduces
background events taken to tape, and the relative efficiency is
straightforward to measure with the events triggered on a coincidence
of 1 or more hits in each set of counters.

For events with much lower multiplicities and/or lower energies, both
the multiplicity and rapidity reach are substantially smaller.  This
strongly impacts the efficiency of the paddle counters, and thus
potentially distorts the centrality estimation.  For these, PHOBOS
uses the full distribution of multiplicities measured in several
$\eta$ regions of the Octagon and Ring multiplicity counters, and
matches them to the distributions measured in a MC simulation,
typically HIJING.  Once the overall multiplicity scale is fit, the
difference in the integrals between data and MC gives a reasonable
estimate of the fraction of observed total cross section.

\subsubsection{STAR}
%--- mike_begin

STAR defines centrality classes for Cu+Cu and Au+Au (d+Au) using
charged particle tracks reconstructed in the TPC (FTPC) over full
azimuth and $|\eta|<0.5$ ($2.5<\eta<4$).  Background
events are removed by requiring the reconstruction of a primary vertex
in addition to either a coincident ZDC (130/200 GeV Au+Au/Cu+Cu) or
BBC (62.4 GeV Au+Au/Cu+Cu) signal.  Vertex reconstruction inefficiency
in low multiplicity events reduces the fraction of the total measured
cross section to, e.g., $(95 \pm 5)$\% for 130 GeV Au+Au.  For MB
events, centrality is defined offline by binning the measured
$\mathrm{d}N_\mathrm{evt} / \mathrm{d}N_\mathrm{ch}$ distribution by
fraction of total cross section.  Glauber calculations are performed
using a Monte Carlo calculation.  STAR enhances central events via an
online trigger using a coincidence between the MB ZDC condition and
large energy deposition in the Central Trigger Barrel (CTB), a set of
240 scintillating slats covering full azimuth and $-1<\eta<1$.  After
offline cuts the central trigger corresponds to the 0-5\% most central
class of events.  STAR has several methods of extracting mean values of Glauber quantities.

(1) STAR reports little dependence on the mean values of
$N_\mathrm{part}$ and $N_\mathrm{coll}$ extracted via the
aforementioned mapping procedure when vastly different models of
particle production were used to simulate the charged particle
multiplicity.  Thus, for many analyses (62.4/130/200 GeV Au+Au) STAR
bypasses simulation of the multiplicity distribution and
instead defines centrality bins from the Monte Carlo calculated
$\mathrm{d}\sigma/\mathrm{d}N_\mathrm{part}$ and
$\mathrm{d}\sigma/\mathrm{d}N_\mathrm{coll}$ distributions themselves
\cite{Adler:2002xw, Adams:2004zg, Adams:2003kv}. Mean values of
Glauber quantities are extracted by binning the calculated
distribution (e.g., $\mathrm{d}\sigma/\mathrm{d}N_\mathrm{part}$)
analogously to the measured
$\mathrm{d}N_\mathrm{evt}/\mathrm{d}N_\mathrm{ch}$.  Potential biases
due to lack of fluctuations in simulated particle production were
evaluated and found to be negligible for all but the most peripheral
events.  Further uncertainties in the extraction of $\langle
N_\mathrm{part} \rangle$ and $\langle N_\mathrm{coll} \rangle$ are
detailed in reference \cite{Adler:2002xw, Adams:2004zg} and are
dominated by uncertainty in $\sigma_\mathrm{inel}^\mathrm{NN}$, the
Woods Saxon parameters of Au and Cu, and the 5\% uncertainty in the
measured cross section.
% The latter is propagated by varying the centrality bin definitions
% within the 5\% uncertainty and recalculating $\langle
% N_\mathrm{part} \rangle$ and $\langle N_\mathrm{coll} \rangle$.

(2) For Cu+Cu and recent studies of elliptic flow fluctuations in
Au+Au, STAR has invoked a full simulation of the TPC multiplicity
distribution \cite{Sorensen:2006nw}, analogous to that performed for
previous d+Au studies described below.

% A suite of triggers used to select rare high $p_T$ events ($\gamma$,
% $\pi^0$, $e$, jets, $J/\Psi$, $\Upsilon$) are beyond the scope of
% this review.
(3) For d+Au events, centrality was defined by both measuring and
simulating the charged particle multiplicity in the FTPC in the
direction of the initial Au beam. The simulated distribution was
constructed using a Monte Carlo Glauber model coupled
to a random sampling of a NBD.  The
NBD parameters were taken directly from measurements of UA5
collaboration at the same rapidity and energy \cite{Ansorge:1988kn}.
For each Monte Carlo event, the NBD was randomly sampled
$N_\mathrm{part}$ times.  After accounting for tracking efficiency,
the simulated $N_\mathrm{ch}$ distribution was found to be in good
agreement with the data \cite{Adams:2003im}.  The mean values of
various Glauber quantities were then extracted as described in Section
\ref{sec:divide}.  A second class of events was also used, where a
single neutron was tagged in the ZDC in the direction of the initial d
beam.  These ``single-neutron'' events are essentially peripheral p+Au
collisions, and the corresponding FTPC multiplicity is again well
described by the Monte Carlo Glauber based simulation
\cite{Adams:2003im}.

\subsection{Acceptance Biases}
Since centrality is estimated using quantities that vary monotonically
with particle multiplicity, one must be careful to avoid associating
fluctuations of an observable with fluctuations in the geometric
quantities themselves.  This is especially true when estimating the
yield per participant pair, when one is estimating the participants
from the yield itself.  Of course, in heavy ion collisions, the
extraordinarily high multiplicities reduce the effect of
auto-correlation bias, as was estimated by STAR \cite{Adams:2003kv}.
However, the RHIC experiments have found that lower multiplicities and
lower energies are quite challenging.  Estimating the number of
participants in d+Au proved particularly delicate, due to
auto-correlations, which were reduced (in HIJING simulations) by using
large regions in pseudorapidity positioned far forward and backward of
mid-rapidity \cite{Back:2004je}.

\subsection{Estimating Geometric Quantities}     

\subsubsection{Total Geometric Cross Section}
\label{sect:totalgeom}
% --- klaus_begin:geometric_cross_section_v1
%MLM --
\begin{figure}[tbp]
\begin{center}
\includegraphics[width=120mm]{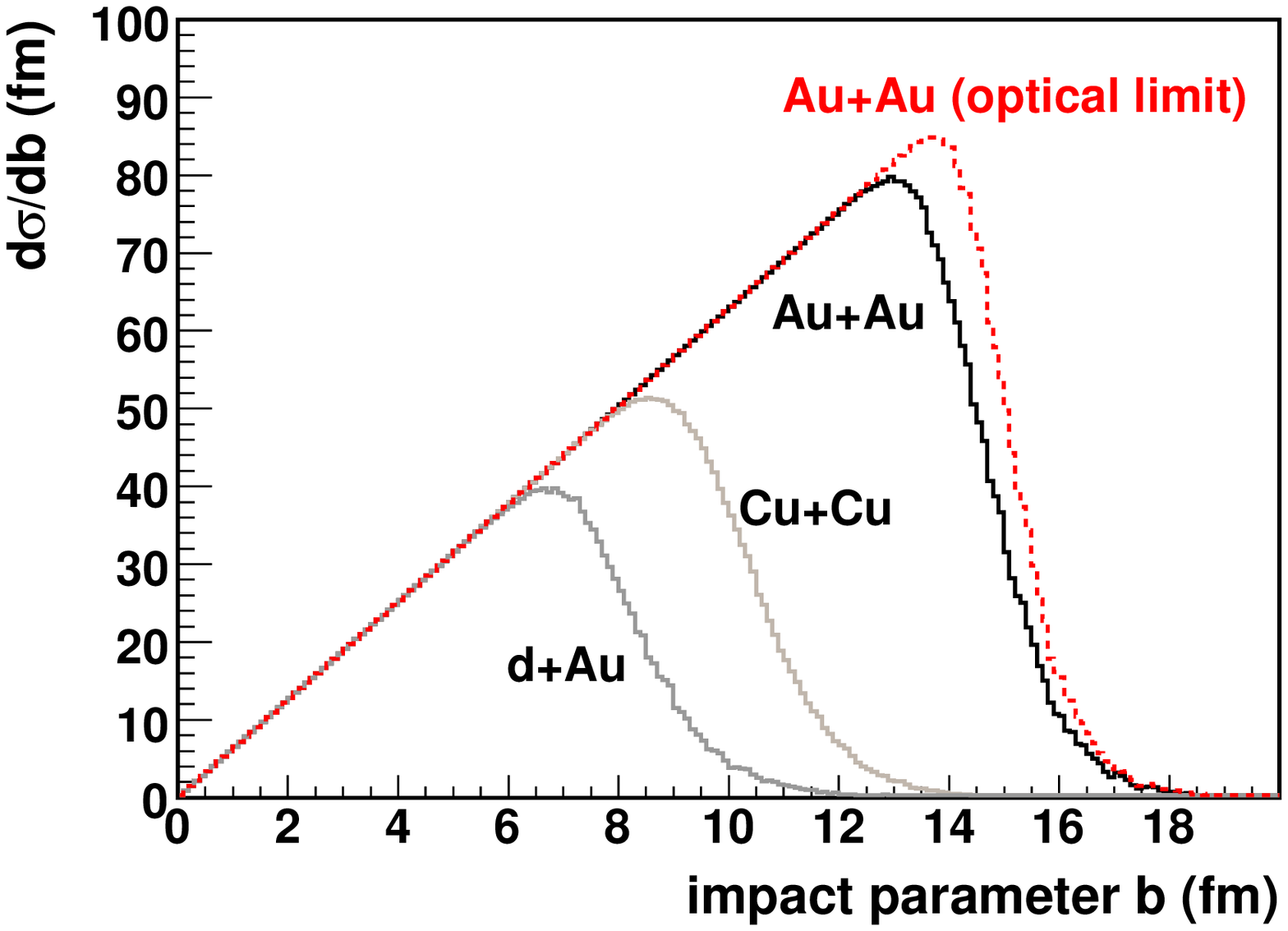}
\caption{ Total geometrical cross section from Glauber Monte Carlo
  calculations (d+Au, Cu+Cu, and Au+Au at $\sqrt{s_\mathrm{NN}} =
  200$~GeV).  The dashed line represents an optical limit calculation
  for Au+Au.
\label{fig:sigma_vs_b}
}
\end{center}
\end{figure}
The total geometric cross section for the collision of two nuclei A
and B, {\it i.e.}, the integral of the distributions
$\mathrm{d}\sigma/\mathrm{d}b$ shown in Figure~\ref{fig:sigma_vs_b}, is
a basic quantity which can be easily calculated in the Glauber Monte
Carlo approach. 
 It corresponds to all Glauber Monte Carlo events with
at least one inelastic nucleon-nucleon collision. In
ultra-relativistic nucleus-nucleus collisions the de Broglie wave
length of the nucleons is small compared to their transverse extent so
that quantum mechanical effects are negligible. Hence, the total
geometric cross section is expected to be a good approximation of the
total inelastic cross section. For the reaction systems in
Figure~\ref{fig:sigma_vs_b} (d+Au, Cu+Cu, and Au+Au at
$\sqrt{s_\mathrm{NN}} = 200$~GeV) the Monte Carlo calculations yield
$\sigma_\mathrm{geo}^\mathrm{d+Au} \approx 2180$~mb,
$\sigma_\mathrm{geo}^\mathrm{Cu+Cu} \approx 3420$~mb, and
$\sigma_\mathrm{geo}^\mathrm{Au+Au} \approx 6840$~mb. The systematic
uncertainties are on the order of $10\,\%$ and are dominated by the
uncertainties of the nuclear density profile.
% --- klaus_end:geometric_cross_section_v1

Also shown in Figure~\ref{fig:sigma_vs_b} is a comparison with anf
optical calculation of $\mathrm{d}\sigma/\mathrm{d}b$, which shows the
effect described in Section \ref{sect:optical}.  Optical limit
calculations do not naturally contain the terms in the
multiple-scattering integral where nucleons ``hide'' behind each
other.  This leads to a slightly larger cross section
($\sigma_\mathrm{geo,optical}^\mathrm{Au+Au} \approx 7280$~mb).  While
this seems like a small perturbation on
$\sigma_\mathrm{geo}^\mathrm{A+B}$, it has a surprisingly large effect
on the extraction of $N_\mathrm{part}$ and $N_\mathrm{coll}$.  This
does not come from any fundamental change in the mapping of impact
parameter onto those variables.  Figure~\ref{fig:dilute_limit} shows the
mean value of $N_\mathrm{coll}$ (upper curve) and $N_\mathrm{part}$
(lower curve) as a function of $b$, where it is seen that the two
track each other very precisely over a large range in impact
parameter, well within the range usually measured by the RHIC
experiments.  The problem comes in when dividing a sample up into bins
in fractional cross section.  While it is straightforward to estimate
the most central bins, one finds a systematic difference of
$N_\mathrm{part}$ between the two calculations as the geometry gets
more peripheral.  
%MLM --
\begin{figure}[tbp]
\begin{center}
\includegraphics[width=100mm]{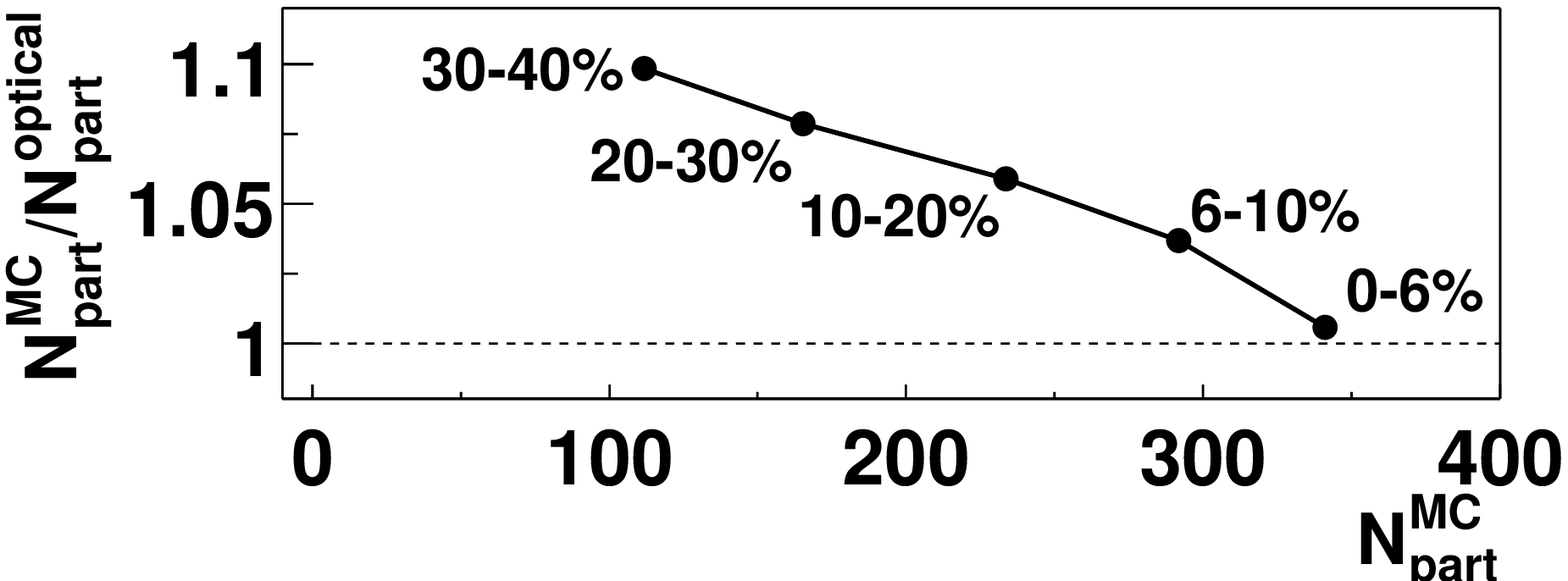}
\caption{ The ratio of $N_\mathrm{part}$ calculated with a Glauber
  Monte Carlo ($N^\mathrm{MC}_\mathrm{part}$) to that calculated with
  an optical approximation ($N^\mathrm{optical}_\mathrm{part}$), for
  the same fraction of the total inelastic Au+Au cross section
  ($\sqrt{s_\mathrm{NN}} = 130$~GeV), plotted as a function of
  $N^\mathrm{MC}_\mathrm{part}$~\cite{Back:2001xy}.
\label{fig:glauber_err}
}
\end{center}
\end{figure}
This is shown in Figure~\ref{fig:glauber_err} for
$\sqrt{s_\mathrm{NN}} = 130$~GeV by comparing the Monte Carlo
calculation in HIJING with the optical limit calculation in
\cite{Kharzeev:2000ph}.

%--- klaus_begin:npart_and_ncoll_v1 
\subsubsection{Participants ($N_\mathrm{part}$) and Binary Collisions ($N_\mathrm{coll}$)}
As described in Section \ref{sect:theory}, the Glauber model is a
multiple collision model which treats a nucleus-nucleus (A+B)
collision as an independent sequence of nucleon-nucleon collisions. A
participating nucleon or wounded nucleon is defined as a nucleon which
undergoes at least one inelastic nucleon-nucleon collision. The
centrality of a A+B collision can be characterized both by the number
of participating nucleons ($N_\mathrm{part}$) and by the number of
binary nucleon-nucleon collisions ($N_\mathrm{coll}$). The average
number of participants $\langle N_\mathrm{part} \rangle$ and
nucleon-nucleon collisions $\langle N_\mathrm{coll} \rangle$ as a
function of the the impact parameter $b$ are shown in
%MLM --
\begin{figure}[tbp]
\begin{center}
\includegraphics[width=120mm]{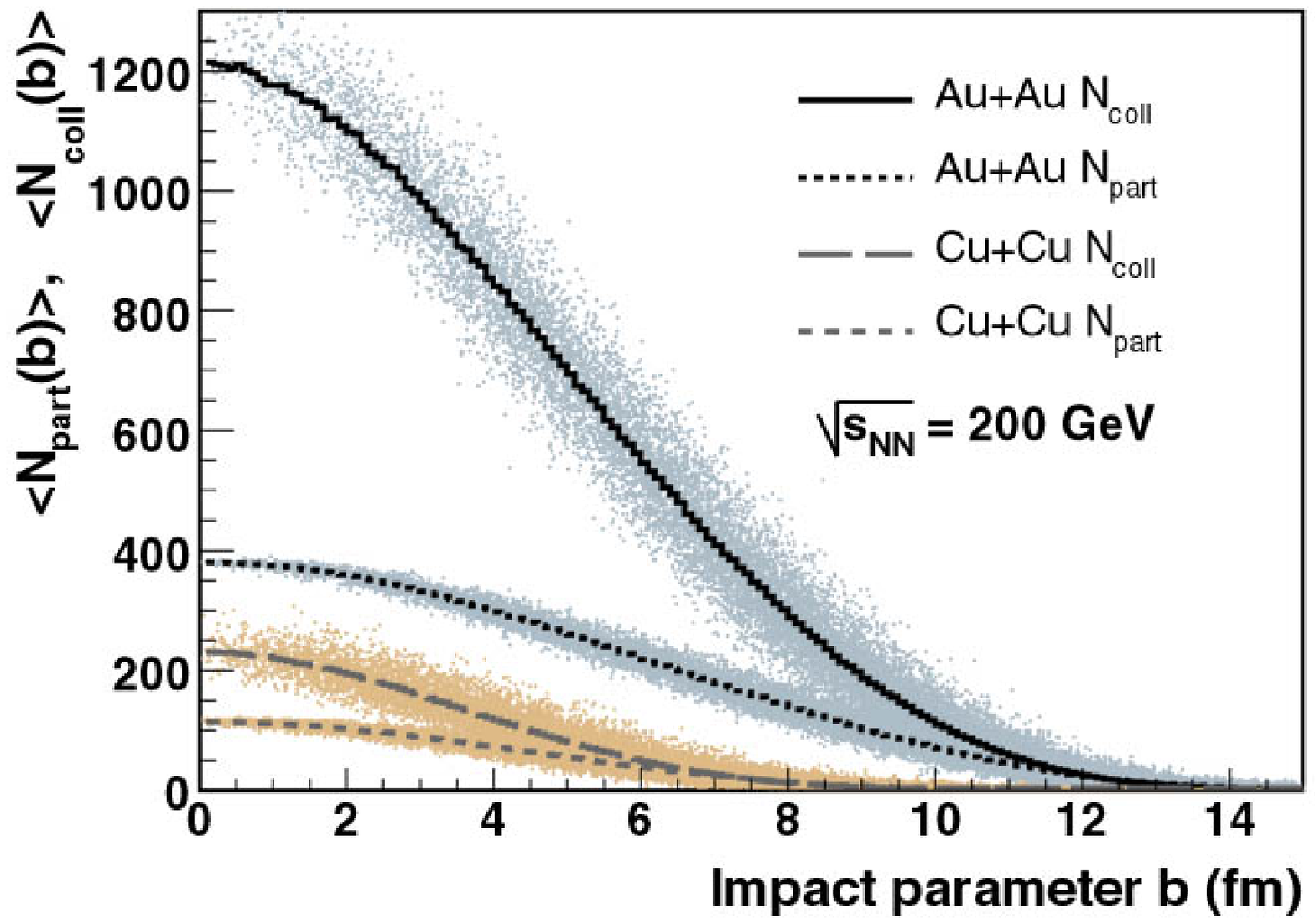}
\caption{ Average number of participants ($\langle N_\mathrm{part}
  \rangle$) and binary nucleon-nucleon collisions ($\langle
  N_\mathrm{coll} \rangle$) along with event-by-event fluctuation of
  these quantities in the Glauber Monte Carlo calculation as a
  function of the impact parameter $b$.
\label{fig:npart_ncoll_vs_b}
}
\end{center}
\end{figure}
Figure~\ref{fig:npart_ncoll_vs_b} for Au+Au and Cu+Cu collisions at
$\sqrt{s_\mathrm{NN}} = 200$~GeV. The event-by-event fluctuations of
these quantities for a fixed impact parameter are illustrated by the
scatter plots.

%MLM --
\begin{figure}[tbp]
\begin{center}
\includegraphics[width=120mm]{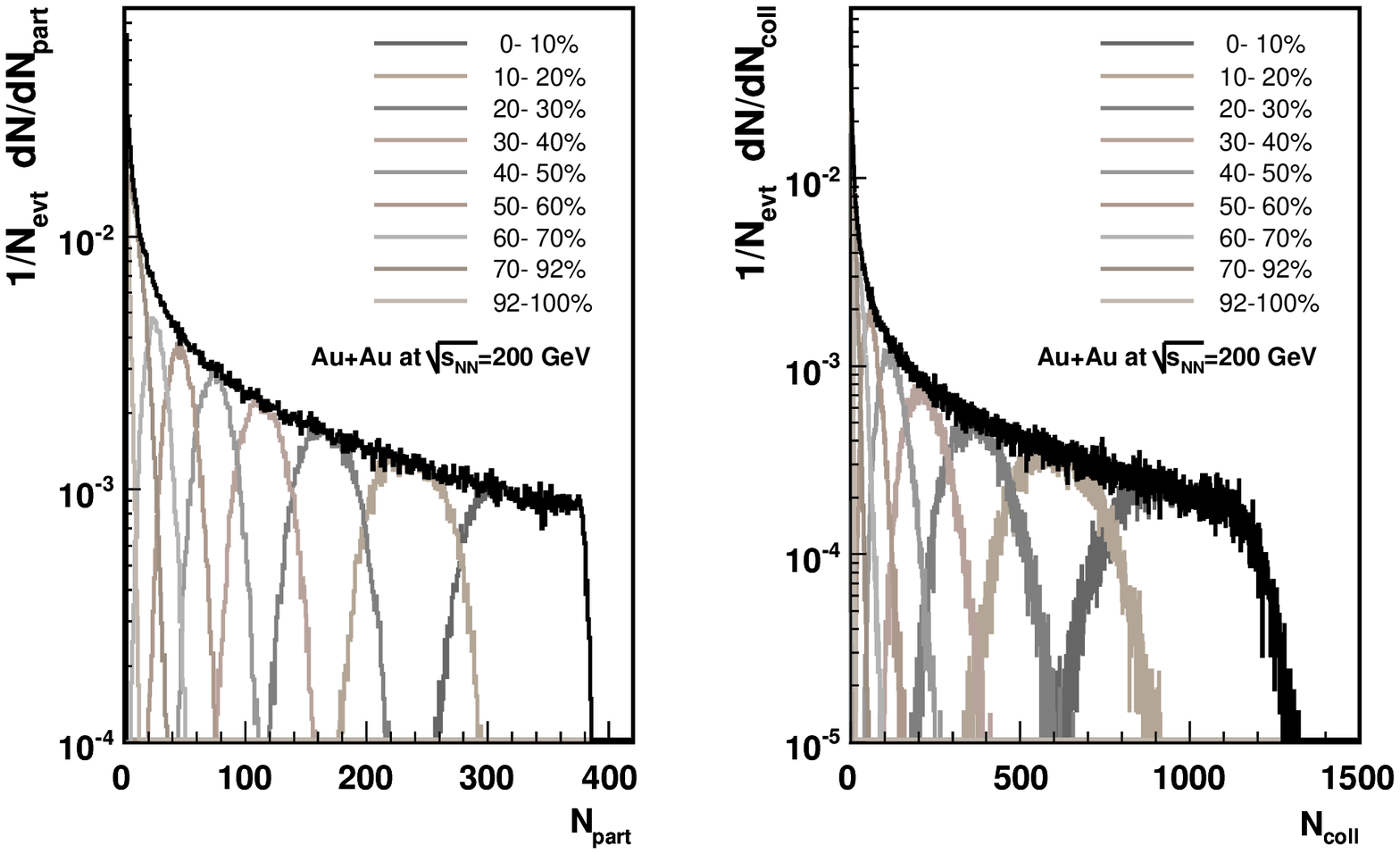}
\caption{
$N_\mathrm{part}$ and $N_\mathrm{coll}$ distribution in Au+Au collisions
at $\sqrt{s_\mathrm{NN}} = 200$~GeV from a Glauber Monte Carlo calculation.
By applying cuts on simulated centrality variables, in this 
case the BBC and ZDC signal as measured by PHENIX, one obtains
$N_\mathrm{part}$ and $N_\mathrm{coll}$ distributions for the
respective centrality class. 
\label{fig:npart_and_ncoll_distr}
}
\end{center}
\end{figure}
The shapes of the $N_\mathrm{part}$ and $N_\mathrm{coll}$
distributions shown in Figure~\ref{fig:npart_and_ncoll_distr} for Au+Au
collisions reflect the fact that peripheral nucleus-nucleus collisions
are more likely than central collisions. $\langle N_\mathrm{part}
\rangle$ and $\langle N_\mathrm{coll} \rangle$ for a given
experimental centrality class, {\it e.g.}, the 10\% most central
collisions, depend on the fluctuations of the centrality variable
which is closely related to the geometrical acceptance of the
respective detector. 
By simulating the fluctuations of the
experimental centrality variable and applying similar centrality cuts
as in the analysis of real data one obtains $N_\mathrm{part}$ and
$N_\mathrm{coll}$ distribution for each centrality class. For
peripheral classes the bias introduced by the inefficiency of the
experimental minimum bias trigger needs to be taken into account by
applying a corresponding trigger threshold on the Glauber Monte Carlo
events. Experimental observables like particle multiplicities can then
be plotted as a function of the mean value of $N_\mathrm{part}$ and
$N_\mathrm{coll}$ distributions.

%MLM --
\begin{figure}[tbp]
\begin{center}
\includegraphics[width=120mm]{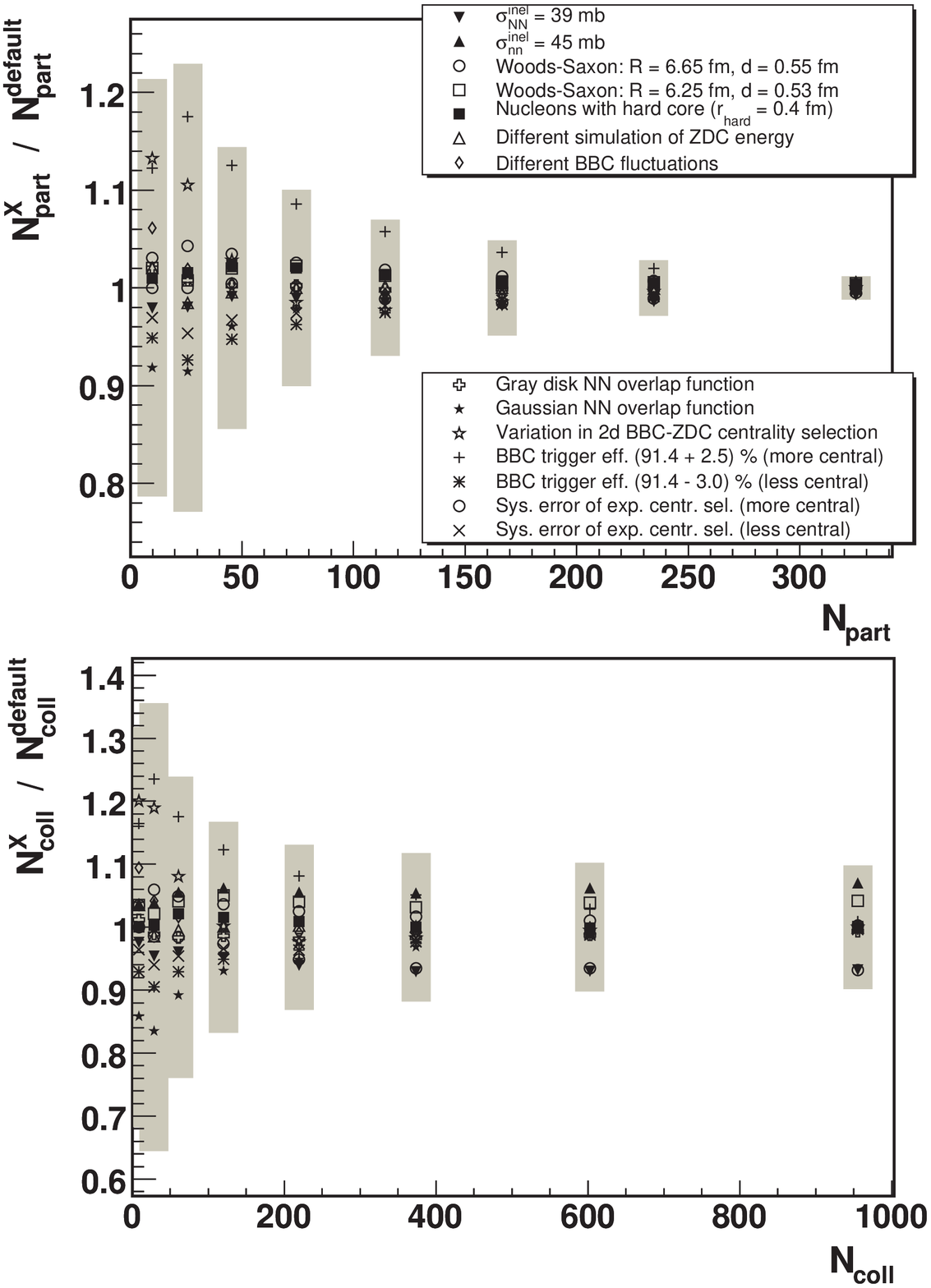}
\caption{Effect of various parameters in
  the Glauber Monte Carlo calculation on $N_\mathrm{part}$ and
  $N_\mathrm{coll}$ for Au+Au collisions at 
  $\sqrt{s_\mathrm{NN}} = 200$~GeV.
\label{fig:syserr_npart_ncoll}
}
\end{center}
\end{figure}
The systematic uncertainties of $N_\mathrm{part}$, $N_\mathrm{coll}$,
and other calculated quantities are estimated by varying various model
parameters. Figure~\ref{fig:syserr_npart_ncoll} shows such a study for
Au+Au collisions at $\sqrt{s_\mathrm{NN}} = 200$~GeV (PHENIX). 
The following effects were considered:
\begin{itemize} 
\item The default value of the nucleon-nucleon cross section of
  $\sigma_\mathrm{NN} = 42$~mb was changed to 39~mb and 45~mb.
\item Woods-Saxon parameters were varied to determine uncertainties
  related to the nuclear density profile.
\item Effects of a nucleon hard core were studied by requiring a
  minimum distance of 0.8~fm between two nucleons of the same nucleus
  without distorting the radial density profile.
\item Parameters of the BBC and ZDC simulation ({\it e.g.} parameters
  describing the finite resolution of these detectors) were varied.
\item The black disk nucleon-nucleon overlap function was replaced by
  ``gray disk'' and Gaussian overlap function \cite{Pi:1992ug} without
  changing the total inelastic nucleon-nucleon cross-section.
\item The origin of the centrality cuts applied in the scatter plot of
  ZDC vs. BBC space was modified in the Glauber calculation.
\item The uncertainty of the efficiency of the minimum bias trigger
  leads to uncertainties as to which percentile of the total inelastic
  cross section actually is selected with certain centrality cuts.
  The centrality cuts applied on the centrality observable simulated
  with the Glauber MC were varied accordingly to study the influence
  on $\langle N_\mathrm{part}\rangle$ and $\langle
  N_\mathrm{coll}\rangle$.
\item Even if the minimum bias trigger efficiency were precisely known
  potential instabilities of the centrality detectors could lead to
  uncertainties as to which percentile of the total cross section is
  selected. This has been studied by comparing the number of events in
  each experimental centrality class for different run periods. The
  effect on $\langle N_\mathrm{part}\rangle$ and $\langle
  N_\mathrm{coll}\rangle$ was again studied by varying the cuts on the
  simulated centrality variable accordingly.
\end{itemize}

The total systematic uncertainty indicated by the shaded boxes in
Figure~\ref{fig:syserr_npart_ncoll} were obtained by adding the
deviations from the default result for each of the items in the above
list in quadrature. The uncertainty of $N_\mathrm{part}$ decreases
from $\sim 20\,\%$ in peripheral collisions to $\sim 3\,\%$ in central
Au+Au collisions. $N_\mathrm{coll}$ has similar uncertainties as
$N_\mathrm{part}$ for peripheral Au+Au collisions. For
$N_\mathrm{part} > 100$ (or $N_\mathrm{coll} > 200$) the relative
systematic uncertainty of $N_\mathrm{coll}$ remains constant at about
$\sim 10\,\%$. Similar estimates for the systematic uncertainties of
$N_\mathrm{part}$ and $N_\mathrm{coll}$ at the CERN SPS energy of
$\sqrt{s_\mathrm{NN}} = 17.2$~GeV were reported in
\cite{Aggarwal:2000bc}.

For the comparison of observables related to hard processes in A+A and
p+p collisions it is advantageous to introduce the nuclear overlap
function $\langle T_\mathrm{AB} \rangle_\mathrm{f}$ for a certain
centrality class f (see section \ref{sect:binary_coll}) which is
calculated in the Glauber Monte Carlo approach as
\begin{equation}
  \langle T_\mathrm{AB} \rangle_\mathrm{f} = 
  \langle N_\mathrm{coll}
  \rangle_\mathrm{f} / \sigma_\mathrm{inel}^\mathrm{NN} \,.
\end{equation}
The uncertainty of the inelastic nucleon-nucleon cross section
$\sigma_\mathrm{inel}^\mathrm{NN}$ doesn't contribute to the
systematic uncertainty. Apart from this $\langle T_\mathrm{AB}
\rangle_\mathrm{f}$ has the same systematics uncertainties as $\langle
N_\mathrm{coll} \rangle_\mathrm{f}$.
%--- klaus_end:npart_and_ncoll_v1 

\subsubsection{Eccentricity}
One of the surprising features of the RHIC data was the strong
event-by-event asymmetries in the azimuthal distributions.  This has
been attributed to the phenomenon of ``elliptic flow'', the
transformation of spatial asymmetries into momentum asymmetries by
means of hydrodynamic evolution.  For any hydrodynamic model to be
appropriate, the system must be sufficiently opaque (where opacity is
the product of density times interaction cross section) such that the
system equilibrates locally at early times.  This suggests that the
relevant geometric quantity for controlling elliptic flow is the
``shape'' of the overlap region, which sets the scale for the
gradients that develop.

The typical variable used to quantify this shape is the 
``eccentricity'', defined as 
\begin{equation}
\epsilon = \frac{\langle Y^2 - X^2 \rangle}{\langle Y^2 + X^2 \rangle} \,.
\end{equation}
Just as with other variables discussed here, there are two ways to
calculate this.  In the optical limit, one performs the averages at a
fixed impact parameter, weighting by either the local participant or
binary collision density.  In the Monte Carlo approach, one simply
calculates the moments of the participants themselves.  Furthermore,
one can calculate these moments with the $X$~axis oriented in two
natural frames.  The first is along the nominal reaction plane
(estimated using spectator nucleons).  The second is along the short
principal axis of the participant distribution itself~\cite{Manly:2005zy}.  
The only mathematical difference
between the two definitions involves the incorporation of the
correlation coefficient $\langle XY \rangle$:
\begin{equation}
\epsilon_\mathrm{part} 
  = \frac{\sqrt{(\sigma^2_x-\sigma^2_y)^2+4(\sigma^2_{xy})^2}}{\sigma^2_x-\sigma^2_y}
\end{equation}

%MLM --
\begin{figure}[tbp]
\begin{center}
\includegraphics[width=90mm]{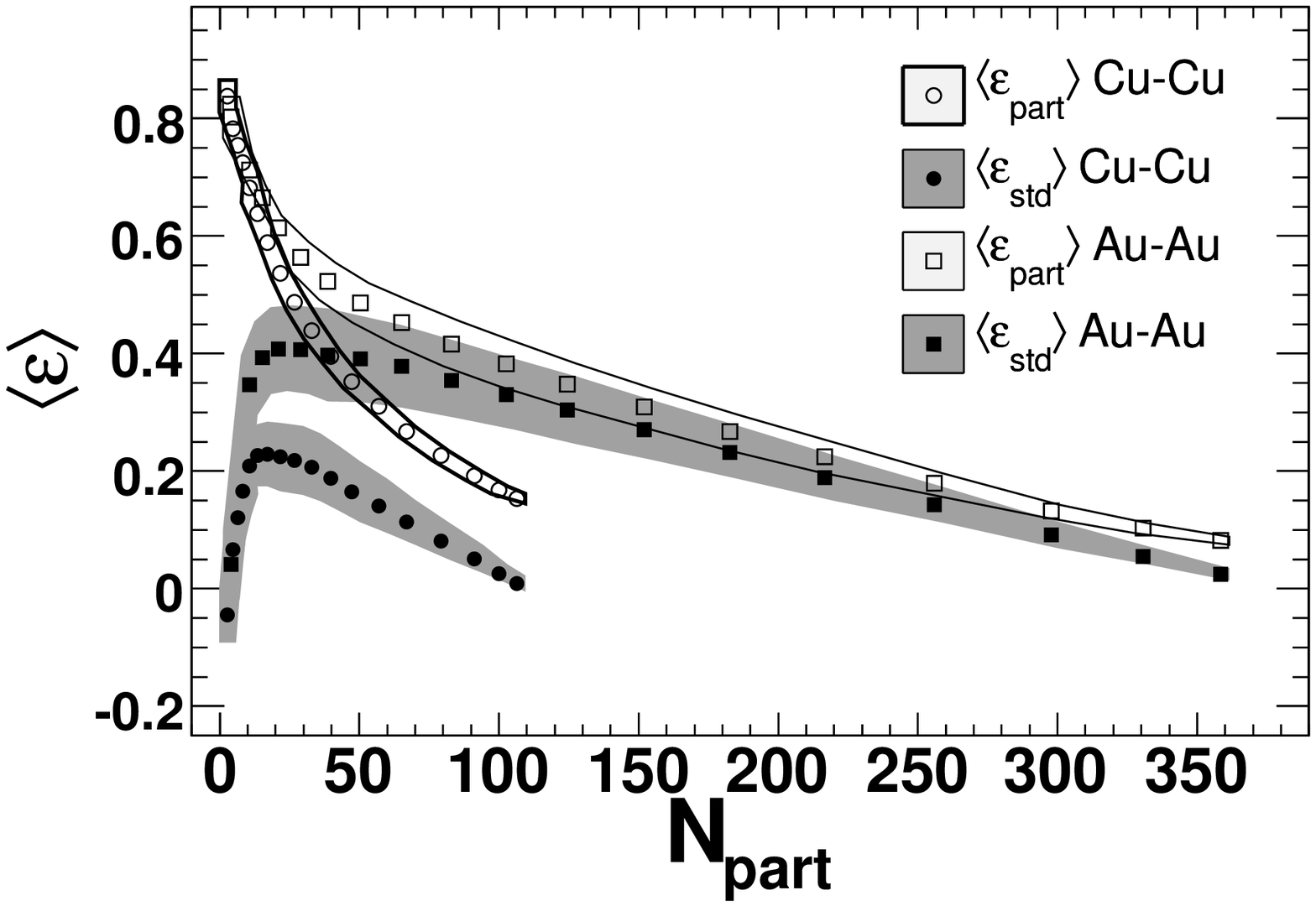}
\caption{
\label{fig:phobos_eccpart_Fig4}
Calculation of eccentricity in Au+Au and Cu+Cu collisions
as a function of $N_\mathrm{part}$.  Both standard and participant eccentricity
are shown.
}
\end{center}
\end{figure}
A comparison of the two definitions is shown in
Figure~\ref{fig:phobos_eccpart_Fig4}.  One sees very different
limiting behavior at very large and small impact parameter.  At large
impact parameter, fluctuations due to small numbers of participants
drive $\epsilon_\mathrm{std} \rightarrow 0$, but
$\epsilon_\mathrm{part} \rightarrow 1$.  As $b=0$,
$\epsilon_\mathrm{std}$ also goes to zero as the system becomes
radially symmetric while $\epsilon_\mathrm{part}$ now picks up the
fluctuations and remains finite.  The relevance of these two
quantities to actual data will be discussed in sections \ref{sec:ecc}
and \ref{sec:ecc_fluct}.

\clearpage

\newpage
\section{Geometric aspects of p+A and A+A phenomena}
\label{sect:physics}
\subsection{Inclusive Charged-Particle Yields (total and mid-rapidity)}

The total multiplicity in hadronic reactions is a measure
of the degrees of freedom released in the collision. 
In the 1970s it was found that the total number of particles
produced in proton-nucleus ($p+A$) collisions was proportional
to the number of participants, i.e. $N_\mathrm{ch} \propto N_\mathrm{part}
= \nu + 1$, where $\nu$ is defined as the number of struck
nucleons in the nucleus~\cite{Elias:1978ft}.  
This experimental fact was instrumental
in establishing $N_\mathrm{part}$ as a fundamental physical variable.
The situation became more interesting when the total multiplicity
was measured in Au+Au at 4 RHIC energies, spanning an order
of magnitude in $\sqrt{s_\mathrm{NN}}$, and was found to be approximately
proportional to $N_\mathrm{part}$ there as well.  
%MLM --
\begin{figure}[tbp]
\begin{center}
\includegraphics[width=80mm]{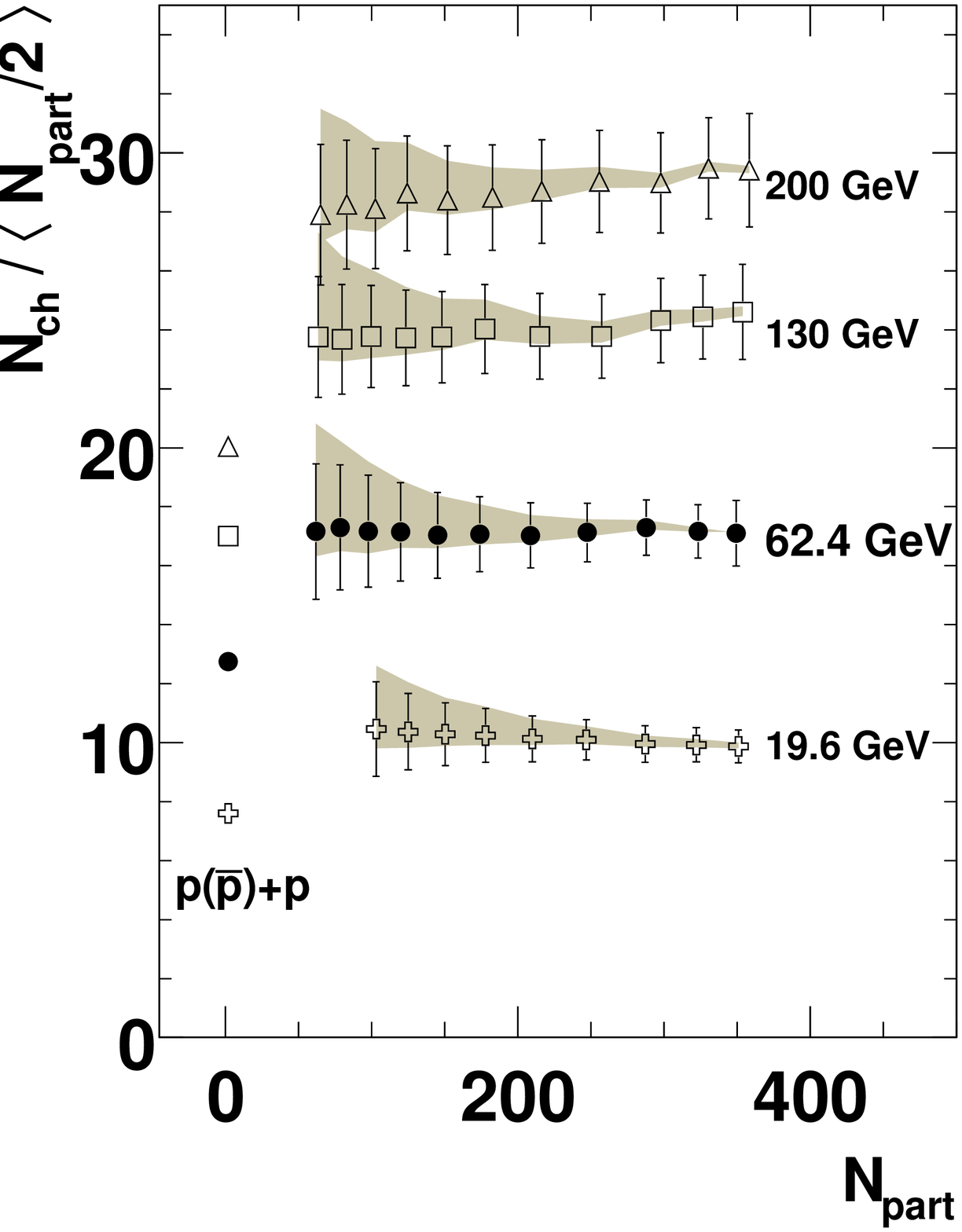}
\caption{
\label{fig:ntot_20_63_130_200}
Total inclusive charged-particle multiplicity ($N_\mathrm{ch}$) divided
by $N_\mathrm{part}/2$ from PHOBOS data at four RHIC energies.  The
data is compared with $p+p$ data or interpolations to unmeasured
energies at $N_\mathrm{part}=2$.
}
\end{center}
\end{figure}
This is shown
in Fig.\ref{fig:ntot_20_63_130_200} with PHOBOS data from
Refs.~\cite{Back:2005hs,Back:2006yw} and is striking if one
considers the variety of physics processes that should contribute to
the total multiplicity.

By contrast, the inclusive charged-particle density near
mid-rapidity ($dN/d\eta(|\eta|<1)$) does {\it not} scale
linearly with $N_\mathrm{part}/2$.  
%MLM --
\begin{figure}[tbp]
\begin{center}
\includegraphics[width=120mm]{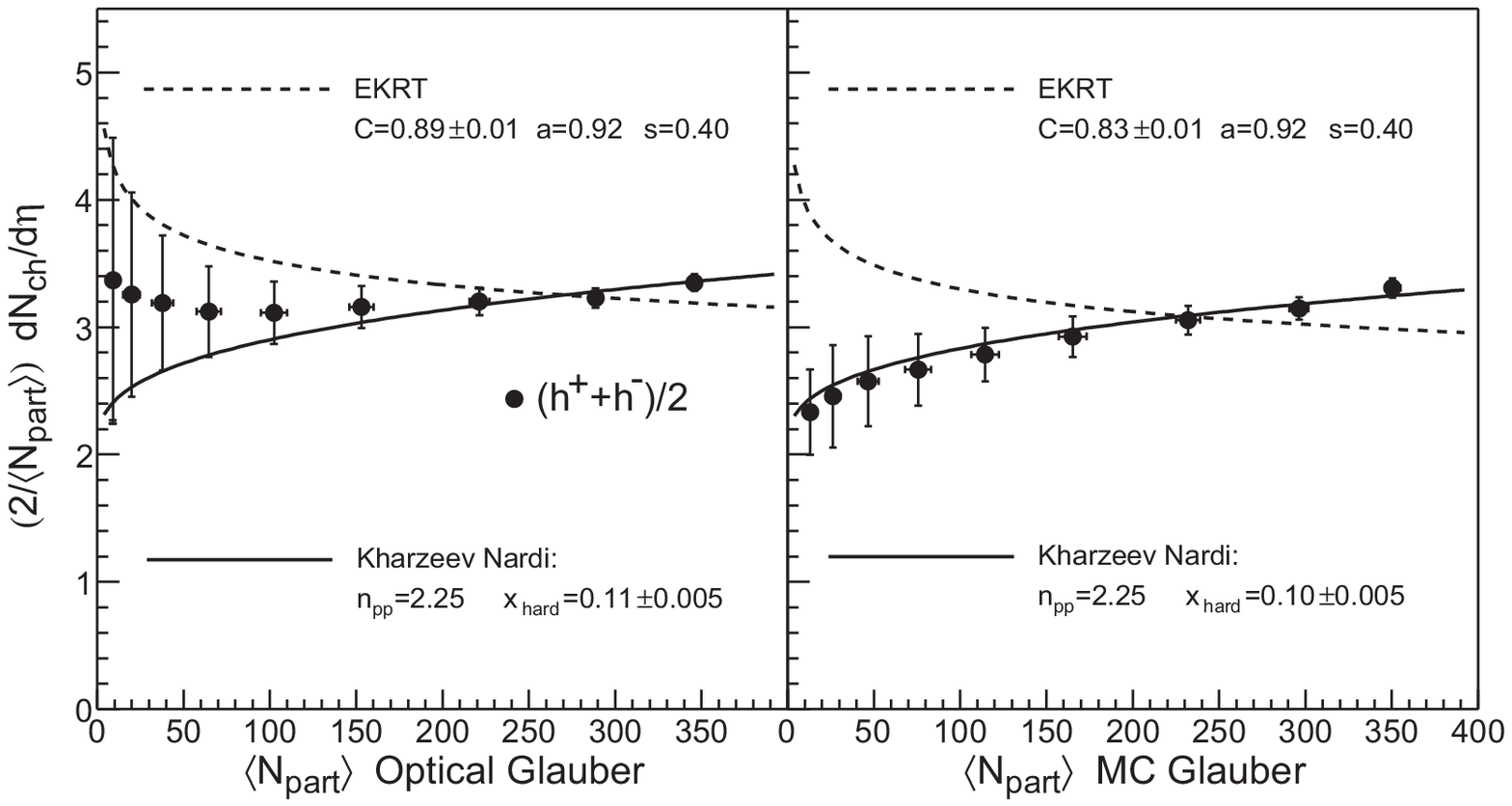}
\caption{
\label{fig:midrap}
Inclusive charged particle multiplicity near midrapidity ($dN/d\eta(|\eta|<1)$)
divided by the number of participating nucleon pairs ($N_\mathrm{part}/2$) estimated 
using the optical approximation (left) and a Glauber Monte Carlo (right)
from STAR.  The data is compared with two-component fits and 
a calculation based on parton saturation.  
}
\end{center}
\end{figure}
This is shown in Fig.~\ref{fig:midrap}
with STAR data, with $N_\mathrm{part}$ estimated both from an optical
calculation (left) as well as a Glauber 
Monte Carlo (right)~\cite{Adams:2003yh}.  
The comparison shows why care must be taken in the estimation
of $N_\mathrm{part}$, since using one or the other gives better agreement
with very different models.  The saturation model of Eskola et al
\cite{Eskola:1999fc} does show scaling with $N_\mathrm{part}$ and agrees
with the data if $N_\mathrm{part}$ is estimated with an optical calculation.
However, it disagrees with the data when the GMC approach is used.
Better agreement with the data can be found with a so-called
``two component'' model, e.g. Ref.~\cite{Kharzeev:2000ph}:
\begin{equation}
\frac{dN}{d\eta} = n_{pp} \left[ (1-x)N_\mathrm{part} + xN_\mathrm{coll} \right]
\end{equation}
This model can be fit to the data by a judicious choice of $x$,
the parameter which controls the admixture of ``hard'' particle 
production.  However there is no evidence of any energy dependence to this
parameter~\cite{Back:2002uc,Back:2004dy} from $\sqrt{s_{NN}} = $ 19.6 to 200 
GeV, suggesting that the source of this dependence has little to nothing
to do with hard or semi-hard processes at all.

\subsection{Hard Scattering: $T_\mathrm{AB}$ Scaling}
\label{sect:binary_coll}

%--- klaus_begin:binary_scaling_v1
The number of hard processes between point-like constituents of the
nucleons in a nucleus-nucleus collision is proportional to the nuclear
overlap function $T_\mathrm{AB}(b)$ \cite{Eskola:1988yh,
  Eskola:1995zt, Vogt:1999jp, Arleo:2004gn, Jacobs:2004qv,
  Tannenbaum:2006ku}. This follows directly from the factorization
theorem in the theoretical description of hard interactions within
perturbative QCD \cite{Owens:1986mp}. In detail, the average yield for
a hard process with cross section $\sigma_\mathrm{hard}^\mathrm{pp}$
in p+p collisions per encounter of two nuclei A and B with impact
parameter $b$ is given by
\begin{equation}
N_\mathrm{hard}^\mathrm{A+B,enc}(b) =
T_\mathrm{AB}(b)\,\sigma_\mathrm{hard}^\mathrm{pp} \,.
\end{equation}
Here $T_\mathrm{AB}(b)$ is normalized so that $\int T_\mathrm{AB}(b)
\, \mathrm{d}^2b = AB$.  $\sigma_\mathrm{hard}^\mathrm{pp}$ can, {\it
  e.g.}, represent the cross-section for the production of
charm-anticharm quark-pairs ($\bar{c}c$) or high-$p_\mathrm{T}$ direct
photons in proton-proton collisions.
 
At large impact parameter  not every encounter of two nuclei A and
B leads to an inelastic collision. Hence, the average number of hard
processes per inelastic A+B collision is given by
\begin{equation}
N_\mathrm{hard}^\mathrm{A+B}(b) =
\frac{T_\mathrm{AB}(b)}{p_\mathrm{inel}^\mathrm{A+B}(b)} \cdot
\sigma_\mathrm{hard}
\label{eq:yield_per_inel_coll}
\end{equation}
where $p_\mathrm{inel}^\mathrm{A+B}(b)$ is the probability of an
inelastic A+B collision. %MLM for an encounter with impact parameter $b$.
In the optical limit $p_\mathrm{inel}^\mathrm{A+B}(b)$ is given by
(cf.  Eq.~\ref{eq:p_int_AB})
\begin{equation}
p_\mathrm{inel}^\mathrm{A+B}(b) = 1 -
\left(1-\sigma_\mathrm{inel}^\mathrm{NN} \frac{T_\mathrm{AB}(b)}{AB}
\right)^{AB} \,.
\end{equation}
Particle yields at RHIC are usually measured as a function of the
transverse momentum ($p_\mathrm{T}$).  If an invariant cross section
$\mathrm{d}\sigma^\mathrm{pp}/\mathrm{d}p_\mathrm{T}$ for a hard
scattering process which leads to the production of a certain particle
$x$ has been measured in p+p collisions, then according to
Eq.~\ref{eq:yield_per_inel_coll} the invariant multiplicity of $x$ per
inelastic A+B collisions with impact parameter $b$ is given by 
\begin{equation}
\frac{1}{N_\mathrm{inel}^\mathrm{AB}}
\frac{\mathrm{d}N_x^\mathrm{A+B}}{\mathrm{d}p_\mathrm{T}} =
\frac{T_\mathrm{AB}(b)}{p_\mathrm{inel}^\mathrm{A+B}(b)} \; \cdot \;
\frac{\mathrm{d}\sigma_x^\mathrm{pp}}{\mathrm{d}p_\mathrm{T}} \,.
\end{equation}
This baseline expectation is purely based on nuclear geometry and
assumes the absence of any nuclear effects.  In reality one needs to
average over a certain impact parameter distribution. As an example we
consider a centrality class f which corresponds to a fixed impact
parameter range $b_1 \le b \le b_2$.  Taking the average in this range
over the impact parameter distribution (weighting factor
$\mathrm{d}\sigma^\mathrm{AB}/\mathrm{d}b = 2 \pi b
p_\mathrm{inel}^\mathrm{A+B}(b)$) leads to
\begin{equation}
\left. \frac{1}{N_\mathrm{inel}^\mathrm{AB}}
  \frac{\mathrm{d}N_x^\mathrm{A+B}}{\mathrm{d}p_\mathrm{T}}
\right|_\mathrm{f} 
= \langle T_\mathrm{AB} \rangle_\mathrm{f} \; \cdot
\; \frac{\mathrm{d}\sigma_x^\mathrm{pp}}{\mathrm{d}p_\mathrm{T}}
\end{equation}
with 
\begin{equation}
\langle T_\mathrm{AB} \rangle_\mathrm{f} = 
\frac{\int_\mathrm{f} \mathrm{d}^2b \,
T_\mathrm{AB}(b)} {\int_\mathrm{f} \mathrm{d}^2b \,
p_\mathrm{inel}^\mathrm{A+B}(b)} = 
\frac{\int \limits_{b_1}^{b_2}
\mathrm{d}b \, 2\pi b \, T_\mathrm{AB}(b)} 
{\int \limits_{b_1}^{b_2} \mathrm{d}b \, 2\pi
b\, p_\mathrm{inel}^\mathrm{A+B}(b)} \,.
\end{equation}
Averaging over the full impact parameter range ($b_1 =0$, $b_2 =
\infty$) yields $\langle T_\mathrm{AB} \rangle_\mathrm{f} = A B /
\sigma_\mathrm{geo}^{A+B}$.  In the Glauber Monte Carlo approach
$\langle T_\mathrm{AB} \rangle_\mathrm{f}$ for a certain centrality
class is calculated as
\begin{equation}
  \langle T_\mathrm{AB} \rangle_\mathrm{f} 
  = \langle N_\mathrm{coll} \rangle_\mathrm{f} / 
  \sigma_\mathrm{inel}^\mathrm{NN}
\end{equation}
where the averaging is done for all A+B collisions with at least one
inelastic nucleon-nucleon collision and whose simulated centrality
variable belongs to centrality class f.

In order to quantify nuclear effects on particles production in hard
scattering processes the nuclear modification factor
$R_\mathrm{AB}(p_\mathrm{T})$ is defined as
\begin{equation}
R_\mathrm{AB}(p_\mathrm{T}) = 
\frac{(N_\mathrm{inel}^\mathrm{AB})^{-1} \,
  \mathrm{d}N_x^\mathrm{A+B}/\mathrm{d}p_\mathrm{T}}
{\langle T_\mathrm{AB} \rangle_\mathrm{f} \, 
 \mathrm{d}\sigma_x^\mathrm{pp}/\mathrm{d}p_\mathrm{T}} \,.
\end{equation}
At high $p_\mathrm{T}$ ($p_\mathrm{T} \ge 2-3$~GeV/$c$ for hadrons and
$p_\mathrm{T} \ge 4-6$~GeV/$c$ for direct photons) particle production
is expected to be dominated by hard processes such that, in the absence
of nuclear effects, $R_\mathrm{AB}$ should be unity.  Due
to their electromagnetic nature high-$p_\mathrm{T}$ direct photons are
essentially unaffected by the hot and dense medium produced in a
nucleus-nucleus collisions so that they should exhibit $T_\mathrm{AB}$
scaling. 
%MLM --
\begin{figure}[tbp]
\begin{center}
\includegraphics[width=100mm]{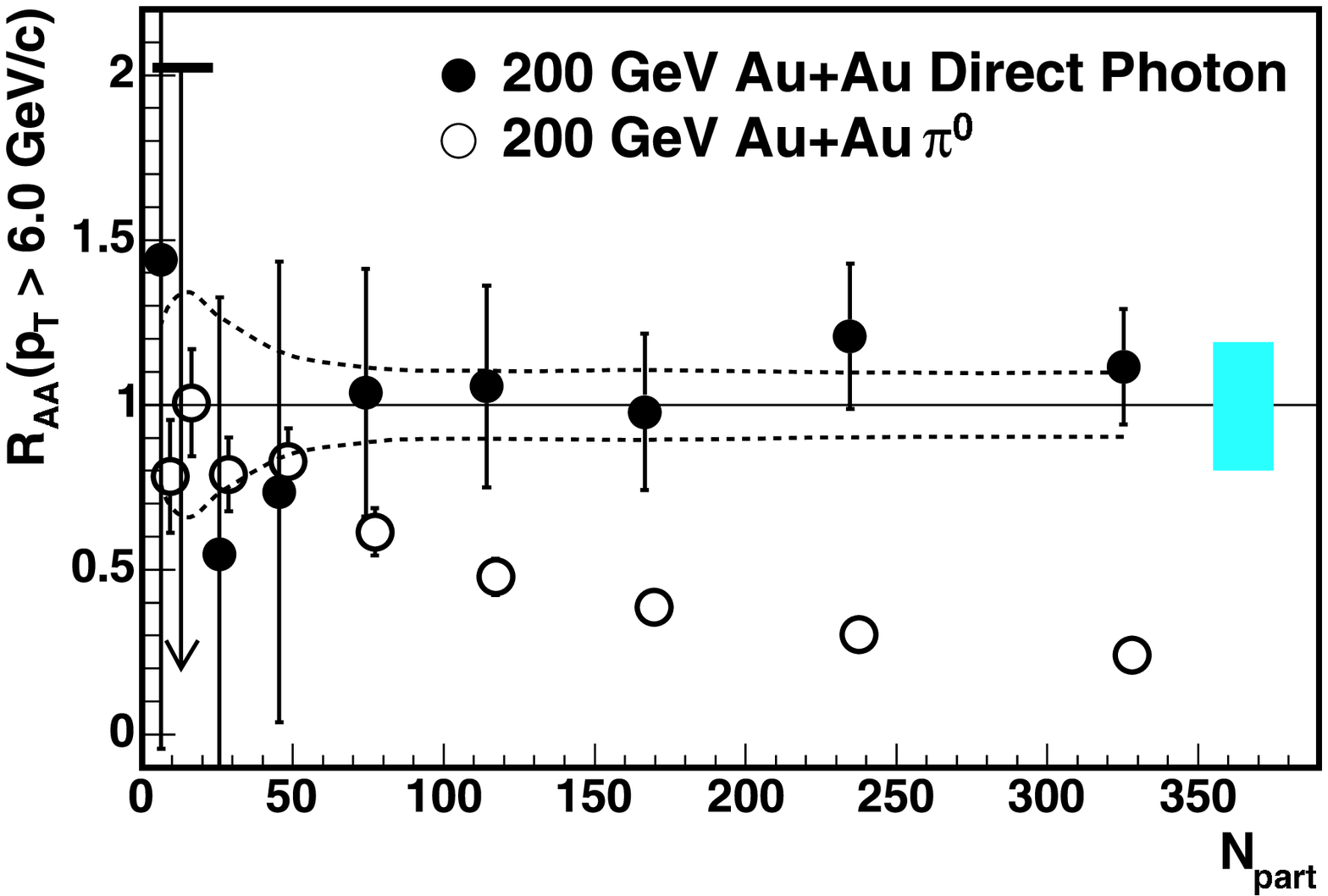}
\caption{
Nuclear modification factor $R_\mathrm{AA}$ in Au+Au collisions 
at $\sqrt{s_\mathrm{NN}} = 200$~GeV for direct-photon and neutral-pion yields
integrated above $p_\mathrm{T} = 6$~GeV/$c$. The dashed lines indicate 
the systematic uncertainties of $\langle T_\mathrm{AB} \rangle_\mathrm{f}$
used in the calculation of $R_\mathrm{AA}$. 
\label{fig:raa_vs_npart}
}
\end{center}
\end{figure}
Fig.~\ref{fig:raa_vs_npart} shows the nuclear modification
factor for direct-photon and neutral-pion yields integrated above
$p_\mathrm{T} = 6$~GeV/$c$ as a function of $N_\mathrm{part}$
\cite{Adler:2005ig}. Direct photons indeed follow $T_\mathrm{AB}$
scaling over the entire centrality range whereas neutral pions are
strongly suppressed in central collisions. This is one of the major
discoveries at RHIC.  The direct-photon measurement is an experimental
proof of $T_\mathrm{AB}$ scaling of hard processes in nucleus-nucleus
collisions. With this observation the most natural explanation for the
suppression of high-$p_\mathrm{T}$ neutral pions is energy loss of
partons from hard scattering in a quark-gluon plasma (jet-quenching)
\cite{Arsene:2004fa, Adcox:2004mh, Back:2004je, Adams:2005dq}.
%--- klaus_end:binary_scaling_v1

\subsection{Eccentricity and relation to Elliptic Flow}
\label{sec:ecc}
Hydrodynamic calculations suggest that spatial asymmetries 
in the initial state are mapped directly into asymmetries
in the final state momentum distribution.  At mid-rapidity,
these asymmetries are manifest in the azimuthal ($\phi$) distributions
of inclusive and identified charged particles, with
the modulation of $\mathrm{d}N/\mathrm{d}\phi \sim 1 + 2 v_2 \cos [2(\phi - \Psi_R)]$
characterized by the $2^{nd}$ Fourier coefficient $v_2$, where $\Psi_R$ defines the angle of the reaction plane for a given event.  It is generally assumed that $v_2$ is proportional to the event eccentricity $\epsilon$,  which was introduced previously.
Glauber calculations are used to estimate the eccentricity, either for an ensemble of events or on an event-by-event basis.

For much of the RHIC program, both calculations were typically carried
out in the ``standard'' reference frame, with the X-axis oriented
along the reaction plane.  Using this calculation method was
apparently sufficient to compare hydrodynamic calculations with
Au+Au data.
However, it was always noticed that the most central events, which
should trend to $\epsilon = 0$ tended to have a significant
$v_2$ value.  This led to the study of the ``participant eccentricity'',
calculated with the X axis oriented along the short principal axis
of the approximately-elliptical distribution of participants in
a Monte Carlo approach~\cite{Manly:2005zy}, described above. 

%MLM --
\begin{figure}[tbp]
\begin{center}
\includegraphics[width=90mm]{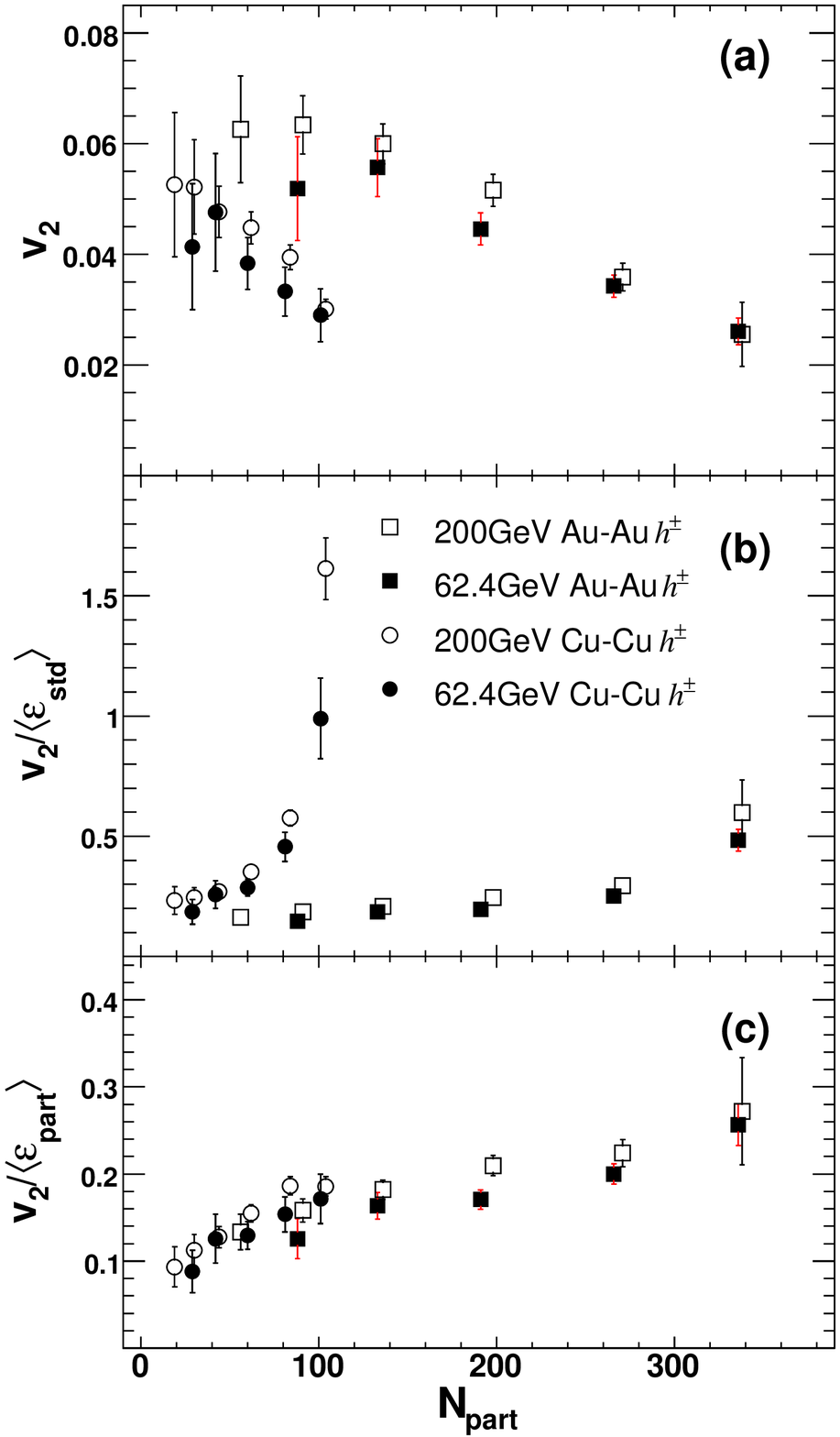}
\caption{
\label{fig:phobos_eccpart_Fig5}
(a) PHOBOS Measurements of $v_2$ for 62.4 and 200 GeV Cu+Cu and Au+Au 
collisions (four system/energy combinations in all, 
from Ref.~\cite{Alver:2006wh}), (b) $v_2$ divided by standard eccentricity and (c) $v_2$ divided by participant eccentricity, showing 
an approximate scaling for all the system/energy combinations.
}
\end{center}
\end{figure}
Fig.~\ref{fig:phobos_eccpart_Fig5} from Ref.~\cite{Alver:2006wh} 
shows $v_2$, the second
Fourier coefficient ($\langle \cos (2[\phi-\Psi_R]) \rangle$) 
of the inclusive particle yield relative to
the estimated reaction plane angle, as a function of $N_\mathrm{part}$.
In hydrodynamic models, $v_2$ is proportional to the eccentricity,
suggesting that $v_2/\epsilon$ should be a scaling variable.
While the raw values of $v_2$ as a function of $N_\mathrm{part}$ peak
at similar levels in Au+Au and Cu+Cu,
it is found that dividing by the standard eccentricity makes
the two data sets diverge.  However, dividing by $\epsilon_\mathrm{part}$
shows that the two systems have similar $v_2/\epsilon_\mathrm{part}$ at
the same $N_\mathrm{part}$.  This shows that the participant eccentricity, 
a quantity calculated in a simple Glauber Monte Carlo approach,
drives the hydrodynamic
evolution of the system for very different energy and geometries.

%--- mike_begin:fluctuations_v1

\subsection{Eccentricity Fluctuations}
\label{sec:ecc_fluct}
As described above, one of the most spectacular measurements at RHIC was the large value of elliptic flow in Au+Au collisions, suggestive of a ``Perfect Liquid."  After the initial measurement \cite{Ackermann:2000tr}, much attention was given to potential biases that could artificially inflate the extraction of $v_2$ from the data, such as ``non-flow" effects (e.g., correlations from jet fragmentation, resonance decay) and event-by-event fluctuations in $v_2$ itself.  Reference \cite{Miller:2003kd} was one of the first analyses to study the effects of fluctuations on extraction of $v_2$.  Using the assumption $v_2 \sim \epsilon_\mathrm{std}$, fluctuations in $\epsilon_\mathrm{std}$ were studied using a Monte Carlo Glauber calculation and comparing $\langle \epsilon_\mathrm{std}^n \rangle$ to $\langle \epsilon_\mathrm{std} \rangle^n$.  Fluctuations were found to play a significant role, where different methods of extraction (e.g. 2-particle vs. higher order cumulants) gave results differing by as much as a factor of two, with the most significant differences found for the most central (0-5\%) and most peripheral (60-80\%) events classes.

Recently in references \cite{Sorensen:2006nw} and \cite{Loizides:2006}, the STAR and PHOBOS collaborations have reported measurements of not only  the $\langle v_2 \rangle$, but also the r.m.s. width $\sigma_{v_2}$.   
%MLM --
\begin{figure}[tbp]
\begin{center}
$\begin{array}{c@{\hspace{0.05in}}c}
\multicolumn{1}{l}{\mbox{\bf}} &
	\multicolumn{1}{l}{\mbox{\bf}} \\ [-0.53cm]
%\multicolumn{1}{l}{\mbox{\bf (a)}} &
%	\multicolumn{1}{l}{\mbox{\bf (b)}} \\ [-0.53cm]
\epsfxsize=2.6in
\epsffile{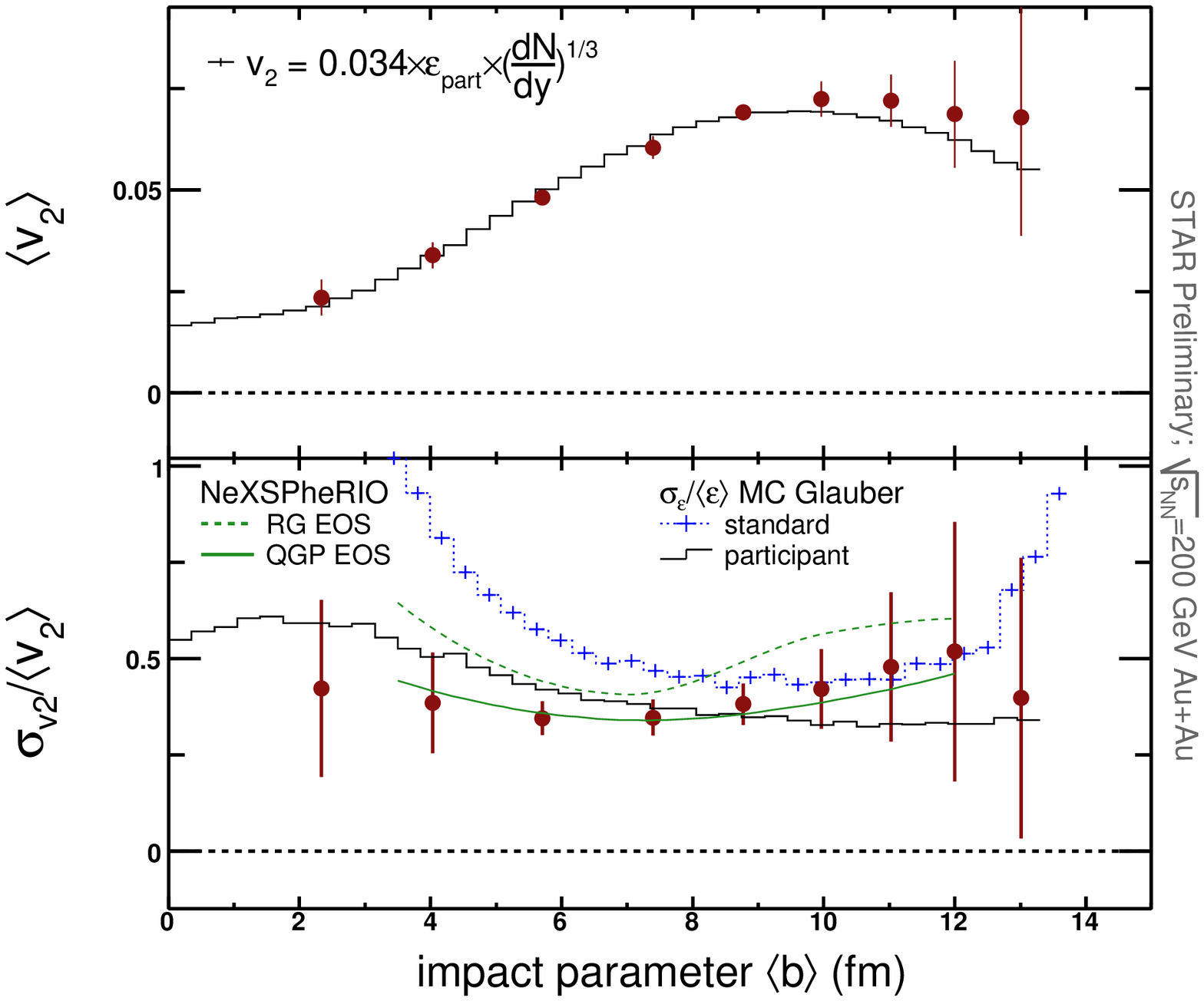} &
	\epsfxsize=2.6in
%	\epsffile{phobos-v2-fluct.eps} \\ [-0.2cm]
	\epsffile{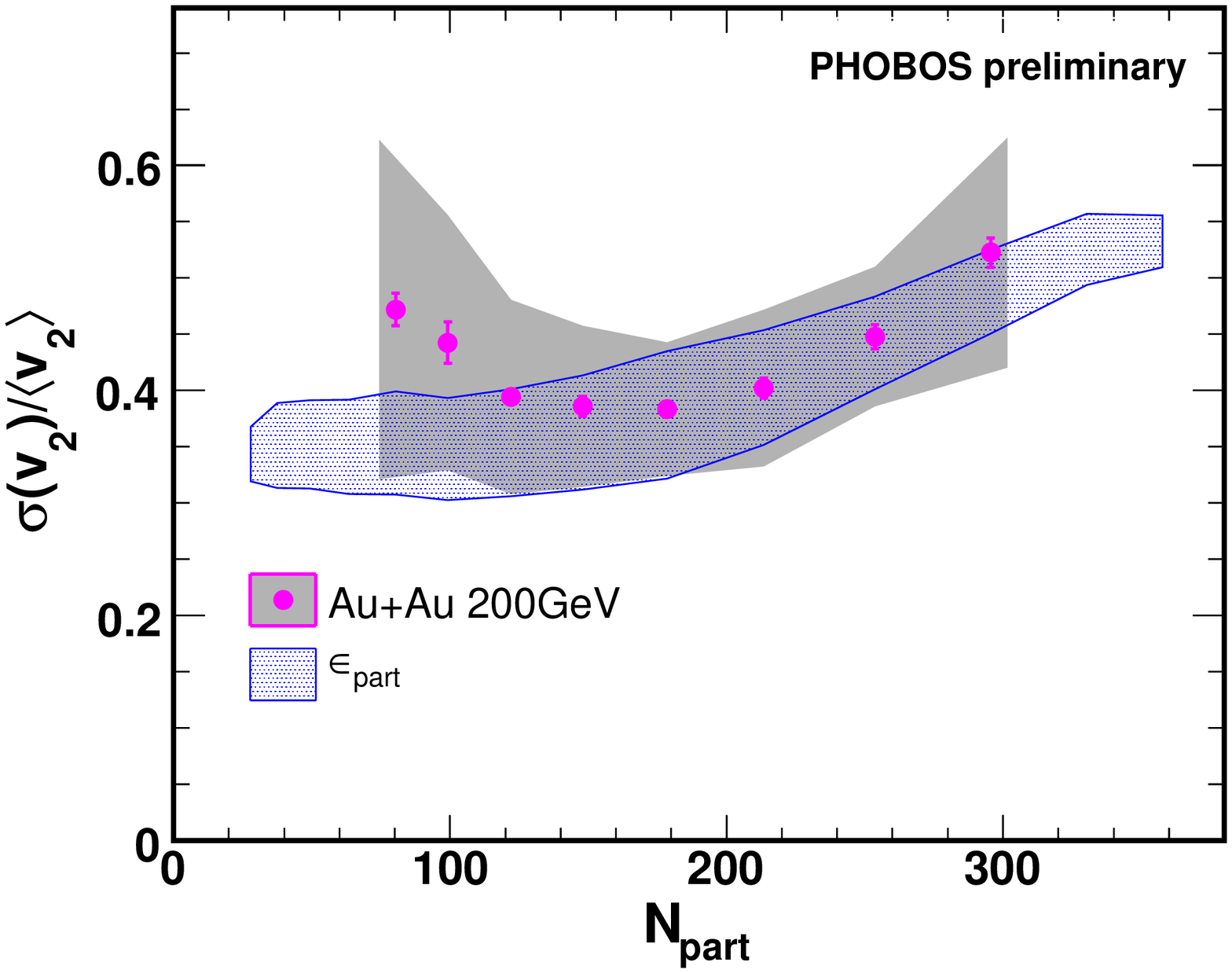} \\ [-0.2cm]
\mbox{\bf (a)} & \mbox{\bf (b)}
\end{array}$
\end{center}
\caption{(a) Elliptic flow mean value (top) and  r.m.s. width scaled by the mean (bottom) as measured by STAR, compared to various dynamical models and a Monte Carlo Glauber calculation \cite{Sorensen:2006nw}.  (b) Gaussian width of $v_2$ divided by mean $v_2$ from PHOBOS~\cite{Loizides:2006}.
}
\label{fig:v2fluct}
\end{figure}
Figure \ref{fig:v2fluct} shows the distribution of  $\sigma_{v_2}/\langle v_2 \rangle$ vs $\langle b\rangle$ from Au+Au data.  The measurements are compared to calculations from various dynamical models as well as Monte Carlo Glauber calculations using both $\sigma_{\epsilon_\mathrm{std}}/\langle \epsilon_\mathrm{std} \rangle$ and $\sigma_{\epsilon_\mathrm{part}}/\langle \epsilon_\mathrm{part} \rangle$.  Clearly the $\epsilon_\mathrm{std}$ description is ruled out while the $\epsilon_\mathrm{part}$ description is in good agreement within the measured uncertainties, implying that the measurements are sensitive to the initial conditions.  The agreement with the $\epsilon_\mathrm{part}$ Glauber calculation further implies that the measured $v_2$ fluctuations are fully accounted for by the fluctuations in the initial geometry, leaving little room for other sources (e.g., Color Glass Condensate).  We note that these analyses are new and the physics conclusions are far from final.  However, this is another excellent example where Glauber calculations are critical in interpreting RHIC data.

%--- mike_end:fluctuations_v1

\subsection{$J/\psi$ absorption in normal nuclear matter}
%--- klaus_begin:jpsi_v1
Due to the large mass of the charm quark $c\bar{c}$ pairs are expected
to be produced only in hard processes in the initial phase of a
nucleus-nucleus collision. The production rate for $c\bar{c}$ pairs is
thus calculable within perturbative QCD which makes them a calibrated
probe of later stages of a heavy ion collision.  In particular, it
was suggested that free color charges in a quark-gluon plasma could
prevent the formation of a $J/\psi$ from the initially produced
$c\bar{c}$ pairs so that $J/\psi$ suppression was initially considered a key
signature of a QGP formation \cite{Matsui:1986dk}. However, a
suppression of $J/\psi$'s relative to the expected production rate for
hard processes was already seen in proton-nucleus (p+A) collisions.
Thus, it became clear that the ``conventional'' $J/\psi$ suppression in p+A
collisions needed to be quantified and extrapolated to A+B collisions
before any conclusions could be drawn about a possible QGP formation
in A+B collisions. Cold nuclear matter effects which affect $J/\psi$
production include the modification of the parton distribution in the
nucleus (shadowing) and the absorption of pre-resonant $c\bar{c}$
pairs \cite{Vogt:1999cu}. For the extrapolation of the latter effect
from p+A to A+A collisions the Glauber model is frequently used \cite{Gerschel:1998zi}, as described in the following.

%MLM --
\begin{figure}[tbp]
\begin{center}
\includegraphics[width=120mm]{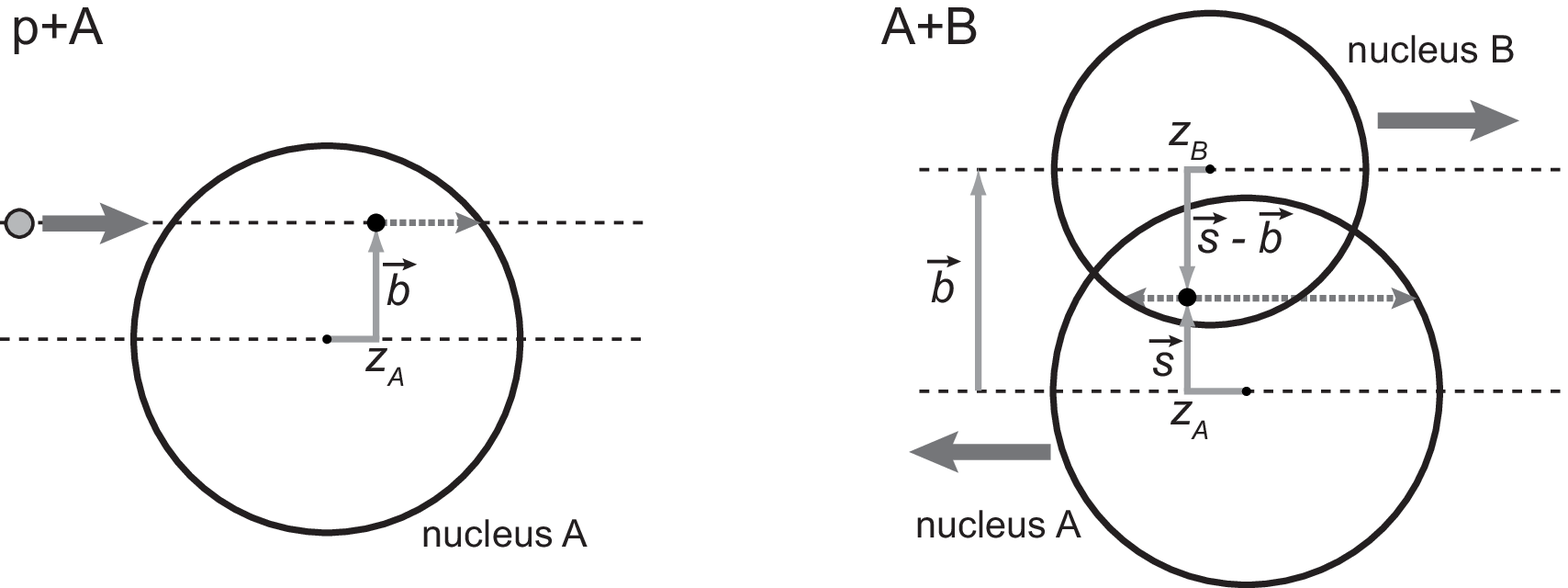}
\caption{Sketch illustrating the calculation of the $J/\psi$ absorption
  in normal nuclear matter in proton-nucleus (left panel) and
  nucleus-nucleus (right panel) collisions). The dashed arrows indicate
  the paths of a $c\bar{c}$ pair created in a hard process.
\label{fig:jpsi}
}
\end{center}
\end{figure}
We first concentrate on p+A collisions (left panel of
Fig.~\ref{fig:jpsi}). Conventional $J/\psi$ suppression is thought to be
related to the path along the $z$ axis in normal nuclear matter a
pre-resonant $c\bar{c}$ pair created at point $({\bf b},z_\mathrm{A})$
needs to travel. Since the $c\bar{c}$ pair is created in a hard
process the location of its production is indeed rather well defined.
Usually the effects of processes which inhibit the formation of a
$J/\psi$ from the pre-resonant $c\bar{c}$ pair are parameterized with
a constant ``absorption" cross section $\sigma_\mathrm{abs}$.
Furthermore, formation time effects are often neglected.  Under these
assumptions the probability for the break-up of a pre-resonant
$c\bar{c}$ state created at $({\bf b},z_\mathrm{A})$ in a collision
with a certain nucleon $j$ of nucleus A then reads \cite{Wong:1995jf,
  Kharzeev:1996yx}
 \begin{equation}
   p_\mathrm{abs}({\bf b},z_\mathrm{A}) = 
   \sigma_\mathrm{abs} \, \hat{T}_\mathrm{A>}({\bf b}, z_\mathrm{A})
   \quad \mathrm{with} \quad
   \hat{T}_\mathrm{A>}({\bf b}, z_\mathrm{A}) = \int \limits_{z_\mathrm{A}}^{\infty}
   \hat{\rho}_\mathrm{A}({\bf b},z) \, \mathrm{d}z \,. 
   \label{eq:p_abs_A}
 \end{equation}
Here $\hat{\rho}_\mathrm{A}$ is the density profile of nucleus A
normalized so that integration over full space yields unity. Hence, the
total survival probability for the $c\bar{c}$ pair is
\begin{equation}
  p_\mathrm{surv}^\mathrm{A}({\bf b},z_\mathrm{A}) =
  \left( 1 - \sigma_\mathrm{abs} \, 
  \hat{T}_\mathrm{A>}({\bf b}, z_\mathrm{A}) \right)^{A-1}
  \approx 
  \exp\left(-(A-1)\, \sigma_\mathrm{abs} \, 
  \hat{T}_\mathrm{A>}({\bf b}, z_\mathrm{A}) \right)
  \label{eq:p_surv_A}
\end{equation}
where the approximation holds for large nuclei $A \gg 1$.  The term
$A-1$ reflects the fact that the $c\bar{c}$-producing nucleon doesn't
contribute to the absorption.  The spatial distribution of the
produced $c\bar{c}$ pairs follows $\hat{\rho}_\mathrm{A}({\bf b},z)$. Thus,
for impact parameter averaged p+A collisions (``minimum bias'') the
expression for the $J/\psi$ absorption in the Glauber model reads
\begin{equation}
  S_\mathrm{p+A} = \frac{\sigma(p+A\rightarrow J/\psi)}
                        {A \cdot \sigma(p+p\rightarrow J/\psi)}
                 = \int \mathrm{d}^2b \, \mathrm{d}z_\mathrm{A} \, 
                   \hat{\rho}_\mathrm{A}({\bf b},z_\mathrm{A}) 
                   \, p_\mathrm{surv}^\mathrm{A}({\bf b},z_\mathrm{A})  \,.
\end{equation}
Fitting the Glauber model expectation to p+A data yields absorption
cross-sections on the order of a few mb at CERN SPS energies
\cite{Alessandro:2003pc}.

As illustrated by the dashed arrows in the right panel of
Fig.~\ref{fig:jpsi}, a $c\bar{c}$ pair created in a collision of two
nuclei A and B has to pass through both nuclei.  Analogous to
Eq.~\ref{eq:p_abs_A} and \ref{eq:p_surv_A} the $J/\psi$ survival
probability for the path in nucleus B can be written as
\begin{equation}
   p_\mathrm{surv}^\mathrm{B}({\bf s}-{\bf b},z_\mathrm{B}) =
   \left( 1 - \sigma_\mathrm{abs} \, 
   \hat{T}_\mathrm{B<}({\bf s}-{\bf b}, z_\mathrm{B}) \right)^{B-1}
   \, , \; 
   \hat{T}_\mathrm{B<}({\bf s}-{\bf b}, z_\mathrm{B}) 
   = \int \limits_{-\infty}^{z_\mathrm{B}}
   \hat{\rho}_\mathrm{B}({\bf s}-{\bf b},z) \, \mathrm{d}z 
   \label{eq:p_surv_B}
\end{equation}
The spatial $c\bar{c}$ production probability density follows
$\hat{\rho}_\mathrm{A}({\bf s},z_\mathrm{A}) \cdot
\hat{\rho}_\mathrm{B}({\bf s}-{\bf b},z_\mathrm{B})$ so that the
normal $J/\psi$ suppression as estimated with the Glauber model for
A+B collisions with fixed impact parameter $b$ is given by
\begin{eqnarray}
\frac{\mathrm{d} S_\mathrm{A+B}}{\mathrm{d}^2b}(b) 
 & = & \frac{1}{A\,B\, \sigma(p+p \rightarrow J/\psi)} \cdot
                 \frac{\mathrm{d}\sigma(AB \rightarrow J/\psi)}{\mathrm{d}^2b} \\ 
 &= & \int \mathrm{d}^2s \, \mathrm{d}z_\mathrm{A} \, \mathrm{d}z_\mathrm{B} \,
      \hat{\rho}_\mathrm{A}({\bf s},z_\mathrm{A})  
      \hat{\rho}_\mathrm{B}({\bf s}-{\bf b},z_\mathrm{B}) \;
      p_\mathrm{surv}^\mathrm{A}({\bf s},z_\mathrm{A}) 
      p_\mathrm{surv}^\mathrm{B}({\bf s}-{\bf b},z_\mathrm{B}) \nonumber
\end{eqnarray}
Here $\mathrm{d}S_\mathrm{A+B}/\mathrm{d}^2b$ is normalized so that
$\int \mathrm{d}^2b \, \mathrm{d}S_\mathrm{A+B}(b)/\mathrm{d}^2b =
1$ for $\sigma_\mathrm{abs} = 0$.  Sometimes the $J/\psi$ suppression
observed in p+A collisions is extrapolated to A+B collisions within
the Glauber framework by calculating an effective path length $L =
L_\mathrm{A} + L_\mathrm{B}$ so that the expected normal $J/\psi$
suppression can be written as \cite{Gerschel:1998zi,
  Alessandro:2003pc, Alessandro:2004ap}
\begin{equation}
S_\mathrm{A+B} = \exp(- L \, \rho_0 \, \sigma_\mathrm{abs}) 
\end{equation}
where $\rho_0 = 0.17$~fm$^{-1}$ is the nucleon density in the center
of heavy nuclei. At the CERN SPS a suppression stronger than expected
from the absorption in cold nuclear matter has been observed in
central Pb+Pb collisions \cite{Alessandro:2004ap}. This so-called
``anomalous'' $J/\psi$ suppression has been discussed as a potential
signal for a QGP formation at the CERN SPS energy.

\clearpage

\newpage
\section{Discussion and the Future}
\label{sect:discussion}
The Glauber model as used in ultra-relativistic heavy-ion physics is
purely based on nuclear geometry. What is left from its origin as a
quantum mechanical multiple scattering theory is the assumption that a
nucleus-nucleus collisions can be viewed as sequence of
nucleon-nucleon collisions and that individual nucleons travel on
straight line trajectories. With the number of participants
$N_\mathrm{part}$ and the number of binary nucleon-nucleon collisions
$N_\mathrm{coll}$ the Glauber model introduces quantities which are
essentially not measurable. Only the forward energy in fixed-target
experiments has a rather direct relation to $N_\mathrm{part}$.

The motivation for the use of these rather theoretical quantities is
manifold. One of the main reasons for the use of geometry-related
quantities like $N_\mathrm{part}$ calculated with the Glauber model is
the possibility to compare centrality dependent observables measured
in different experiments. Moreover, the comparison of different
reaction system as a function of geometric quantities often
leads to new insights. Basically all experiments calculate
$N_\mathrm{part}$ and $N_\mathrm{coll}$ in a similar way using a Monte
Carlo implementation of the Glauber model so that the theoretical bias
introduced in the comparisons is typically small.
Thus, the Glauber model provides a crucial
interface between theory and experiment. 
%Often the centrality
%dependence of theoretically predicted quantities is given in terms of
%$N_\mathrm{part}$ or $N_\mathrm{coll}$. In such comparisons it is
%important to keep in mind that Glauber calculations in the optical
%limit approximation differ from Monte Carlo results in the case of
%peripheral nucleus-nucleus collisions.

The widespread use of the Glauber model is related to the fact that
indeed many aspects of ultra-relativistic nucleus-nucleus collisions
can be understood purely based on geometry. A good example is the
total charged particle multiplicity which scales as $N_\mathrm{part}$
over a wide centrality and center-of-mass energy range. Another
example is the anisotropic momentum distribution of low-$p_\mathrm{T}$
($p_\mathrm{T} < \sim 2$~GeV/$c$) particles with respect to the
reaction plane.  This so-called elliptic flow has its origin in the
spatial anisotropy of the initial overlap volume in non-central
nucleus-nucleus collisions. It is a success of the Glauber model that
event-by-event fluctuations of the spatial anisotropy of the overlap
zone as calculated in the Monte Carlo approach appear to be relevant
for the understanding of the measured elliptic flow.  In this
way, a precise understanding of the Glauber picture has been
of central concern for understanding the matter produced at RHIC
as a ``near perfect fluid''.

The study of particle production in hard scattering processes is
another important field of application for the Glauber model.
According to the QCD factorization theorem the only difference between
p+p and A+A collisions in the perturbative QCD description in the
absence of nuclear effects is the increased parton flux. This
corresponds to a scaling of the invariant particle yields with the
number of binary nucleon-nucleon collisions ($N_\mathrm{coll}$) as
calculated with the Glauber model. The scaling of hard processes with
$N_\mathrm{coll}$ or $T_\mathrm{AB}$ was confirmed by the measurement
of high-$p_\mathrm{T}$ direct photons in Au+Au collisions at RHIC.
This supported the interpretation of a deviation from $T_\mathrm{AB}$
scaling for neutral pions and other hadrons (high-$p_\mathrm{T}$
hadron suppression) as a result of parton energy loss in a quark-gluon
plasma.

Future heavy ion experiments, both at RHIC and at the LHC will further
push our understanding of nuclear geometry.  As RHIC experiments
study more complex multiparticle observables, the understanding of fluctuations
and correlations even in something as apparently simple as the
Glauber Monte Carlo will become a limiting factor in interpreting data.
And as the study of high $p_T$ phenomena involving light and heavy
flavor becomes prominent in the RHIC II era, the understanding of
nuclear geometry, both experimentally and theoretically, will limit
the experimental systematic errors.  At the LHC, the precision of
the geometric calculations will be limited by the knowledge of
$\sigmann$, which should be measured in the first several years
of the $p+p$ program.  After that, Glauber calculations will be
a central part of understanding the baseline physics of heavy
ions at the LHC in terms of nuclear geometry.  It is hoped that
this review will prepare the next generation of relativistic
heavy ion physicists for tackling these issues.

\section{Acknowledgements}
The authors would like to thank our colleagues for illuminating
discussions, especially 
Mark Baker,
Andrzej Bialas,
Wit Busza,
Jamie Dunlop, 
Roy Glauber,
Ulrich Heinz,
Constantin Loizides,
Steve Manly,
Alexander Milov,
Dave Morrison,
Jamie Nagle,
Mike Tannenbaum,
and Thomas Ullrich.
We would like to thank the Editorial staff of Annual Reviews
for their advice and patience.  Miller acknowledges the support of the MIT Pappalardo Fellowship in Physics.  This work was supported in part
by the Office of Nuclear Physics of the U.S. Department of Energy under
contracts: DE-AC02-98CH10886, DE-FG03-96ER40981, DE-FG02-94ER40818.

\end{document}